\documentclass {report}
\usepackage [utf8] {inputenc}
\usepackage [T1] {fontenc}
\usepackage {hyperref}
\usepackage {url}
\usepackage {breakurl}
\usepackage {geometry}

\begin {document}
\thispagestyle {empty}
\ 
\vspace {2cm}

\begin {center}
{\emph {\Large (work in progress)}}

\vspace {2cm}

{\Huge Context-Free Language Theory Formalization}

\vspace {2cm}

{\Large Marcus Vinícius Midena Ramos}

\vspace {0.5cm}

Universidade Federal de Pernambuco (UFPE)\\ Recife, PE, Brazil\\ \texttt {mvmr@cin.ufpe.br}

\vspace {0.5cm}

Universidade Federal do Vale do São Francisco (UNIVASF)\\ Juazeiro, BA, Brazil\\ \texttt {marcus.ramos@univasf.edu.br}

\vspace {2cm}

Ruy J. G. B. de Queiroz \\
\emph {Supervisor} \\
Universidade Federal de Pernambuco (UFPE)\\ 
Recife, PE, Brazil\\
\texttt {ruy@cin.ufpe.br}

\vspace {2cm}
\today
\end {center}

\begin {abstract}
Proof assistants are software-based tools that are used in the mechanization of proof construction and validation in mathematics and computer science, and also in certified program development. Different tools are being increasingly used in order to accelerate and simplify proof checking. Context-free language theory is a well-established area of mathematics, relevant to computer science foundations and technology. This proposal aims at formalizing parts of context-free language theory in the Coq proof assistant. This report presents the underlying theory and general characteristics of proof assistants, including Coq itself, discusses its use in relevant formalization projects, presents the current status of the implementation, addresses related projects and the contributions of this work. The results obtained so far include the formalization of closure properties for context-free grammars (under union, concatenation and closure) and the formalization of grammar simplification. Grammar simplification is a subject of high importance in computer language processing technology as well as in formal language theory, and the formalization refers to the fact that general context-free grammars generate languages that can be also generated by simpler and equivalent context-free grammars. Namely, useless symbol elimination, inaccessible symbol elimination, unit rules elimination and empty rules elimination operations were described and proven correct with respect to the preservation of the language generated by the original grammar.

\vspace {0.5cm}

\end {abstract}

\tableofcontents

\chapter {Introduction}

The fundamental mathematical activity of stating and proving theorems has been traditionally done by professionals that rely purely on their own personal efforts in order to accept or refuse any new proposal, after extensive manual checking. This style of work, which has been used for centuries by mathematicians all over the world, is now changing thanks to computer technology support. 

The so called ``proof assistants'' are software tools that are used in regular computers and offer a friendly and fully checked environment where one can describe mathematical objects and its properties, and then prove theorems about them. The main advantage of their use is to guarantee that proofs are fully checked and correctly constructed. Support for proof automation (that is, automatic proof construction) is also provided in different levels, but this is not the main focus. When applied to program development, these tools are also helpful in checking the correctness of an existing program and also in the construction of correct programs. In order to obtain these benefits, however, one must first be familiar with the underlying mathematical theory of the chosen tool, as well as the associated languages and methodologies required to obtain these proofs. 

Context-free language theory is a fundamental area in mathematics and computer science, which was extensively developed during the 1960s and 1970s. Fundamental to the study and development of computer languages, as well as computer languages processing technology, this theory also leads to important conclusions about the computation process itself. 

The objective of this work is to formalize a substantial part of context-free language theory in the Coq proof assistant, making it possible to reason about it in a fully checked environment, with all the related advantages. Initially, however, the focus has been restricted to context-free grammars and associated results. Stack automata and their relation to context-free grammars shall be considered in the future.

In order to follow this paper, the reader is required to have basic knowledge of Coq and of context-free language theory. For the beginner, the recommended starting points for Coq are the book by Bertot \cite {bertot-2004}, the online book by Pierce \cite {pierce} and a few tutorials avaliable on \cite {coq-site}. More information on the Coq proof assistant, as well as on the syntax and semantics of the following definitions and statements, can be found in \cite {coq-2012}. Background on context-free language theory can be found in \cite {sudkamp-2006}, \cite {hopcroft-1979} or \cite {ramos-2009}, among others.

The motivation for this work comes from (i) the large amount of formalization already existing for regular language theory; (ii) the apparent absence of a similar formalization effort for context-free language theory, at least in the Coq proof assistant and (iii) the high interest in context-free language theory formalization as a result of its practical importance in computer technology (e.g. correctness of language processing software). More information on related works is provided in Chapter \ref {cha-related}.

This report refers to a project whose aim is to represent and formalize main results of context-free language theory in the Coq proof assistant, using its underlying mathematical theory, the Calculus of Inductive Constructions. Despite its huge number of results, fundamental to both theoretical computer science and practical computer technology, this area has not received too much attention in respect to the formalization and development of fully checked demonstrations. As it will become clear in Chapter \ref {cha-related}, only recently some efforts in this direction have appeared, but a lot remains to be done.

The benefits expected with this work are the creation of a library of language theory objects, algorithms, lemmas and theorems, and related machine-checked proofs. As a side effect, the author expects to acquire increased insight in general theory formalization and correct program development.

The present work starts in Chapter \ref {cha-formal} with a brief review of the theoretical background required to understand and use modern proof assistants. Although this technology is recent (first experiments date from late 1960s), the theory behind them dates from the beginning of the 20th century and should be considered independently. The structure of this chapter follows more or less the logical development and interrelationship of its constituents: logic and natural deduction, the lambda calculus of Church (both untyped and typed versions), the Curry-Howard Correspondence, type theory and the Calculus of Inductive Constructions. 

Proof assistants, although available in different forms and flavors, share many characteristics in common. These are described briefly in Chapter \ref {cha-pa}, along with a summary of their development over time.

Among the many proof assistants currently available, Coq has been chosen for the present project. Its main features are then described in Chapter \ref {cha-coq}.

New and different uses of Coq and other proof assistants are announced frequently. The acceptance, usage and importance of proof assistants in computer formalizations and machine-checked developments are increasing very rapidly. Also, the complexity of the projects that are being addressed by such tools. This can be confirmed by the sampler of projects (mostly in Coq) presented and discussed in Chapter \ref {cha-sampler}. These include, for example, the proof of the Four Color Theorem by Georges Gonthier and Benjamin Werner at Microsoft Research in 2005 \cite {gonthier-four} and also the demonstration of the Feit-Thompson Theorem by a group led by Georges Gonthier in 2012 \cite {gonthier-feit}. Also, there are important projects in the areas of mathematics \cite {arXiv-hales}, compiler certification \cite {leroy-2009} and digital security certification \cite {javacard}, among others \cite {coq-projects-lang} \cite {coq-projects-math}. 

Chapter \ref {cha-lang} introduces the application area for this proposal, namely context-free language theory. 

The general idea of mechanizing context-free language theory in the Coq proof assistant is then discussed in Chapter \ref {cha-current}. In particular, we present the strategy adopted, the goals that are being set and a few results obtained so far.  The formalization of closure properties is then discussed in details in Chapter {closure}, and the formalization of grammar simplification in Chapter {simplification}.

General language and automata theory formalization is not a completely new area of research. Some papers have already been published and some results, usually limited to a few related representations, have already been obtained. In Chapter \ref {cha-related}, a summary of these accomplishments is presented, which are then considered in the context of our own proposal.

Finally, the main contributions expected as a consequence of this work are presented and discussed in Chapter \ref {cha-contrib}. Final conclusions are then presented in Chapter \ref {cha-conclusions}.	

\chapter {Formal Mathematics}
\label {cha-formal}
Common sense has it that mathematics is \emph {per se} a formal system, with formal (i.e. complete and non-ambiguous) languages to describe its entities and properties, as well as formal reasoning and computation rules that leave no room for misinterpretation nor allow for inconsistencies.

This is not always true, however. Indeed, most of mathematics that has been developed up to the present days still relies on informal notations and reasoning/computation rules that have a negative impact in both human-human communication and, recently, also restrict human-machine communication. Besides this, as more and more is learned about mathematics and more widespread is its use in virtually all areas of human knowledge, the length and complexity of demonstrations is increasing, making it a very difficult task for a human being to develop and check a proof with its own resources and capabilities alone.

As noted by John Harrison in \cite {harrison-1996}, 

\begin {quote}
``Mathematics is generally regarded as the exact subject par excellence. But the language commonly used by mathematicians (in papers, monographs, and even textbooks) can be remarkably vague; perhaps not when compared with everyday speech but certainly when compared with the language used by practitioners of other intellectual disciplines such as the natural sciences or philosophy... \\
The formalization of mathematics addresses both these questions, precision and correctness. By formalization we mean expressing mathematics, both statements and proofs, in a (usually small and simple) formal language with strict rules of grammar and unambiguous semantics. The latter point bears repeating: just writing something in a symbolic language is not sufficient for formality if the semantics of that language is ill-defined... \\
In formalizing mathematics, we must rephrase informal constructs in terms of formal ones, which is merely an extreme case of defining non-rigorous concepts rigorously.''
\end {quote}

For this reason, formal mathematics has emerged as an area of knowledge whose objective is to establish simple, precise and non-ambiguous reasoning and computation rules that can be used by humans, with the aid of computers, in order to develop and check proofs with higher confidence, in shorter periods of time, thus increasing our productivity and capability of dealing with higher complexity matters. Formalization is, according to Freek Wiedijk, ``computer encoded mathematics'' \cite {wiedijk-2008}. A formal proof is, in John Harrison's words, ``a proof written in a precise artificial language that admits only a fixed repertoire of stylized steps. This formal language is usually designed so that there is a purely mechanical process by which the correctness of a proof in the language can be verified'' \cite {harrison-2008}. 

Proof assistants are, thus, the tools that enable the construction of formal proofs, expressed in a specialized formal language, and the verification of their correctness.

Wiedijk explains why he considers formalization as one of the three most important revolutions that mathematics has undergone in its whole history, along with the introduction of the concepts of \emph {proof} by the greeks in the 4th century and of \emph {rigor} in the 19th century, and provides a synthetic yet strong expression of its importance in our times. In his own words \cite {wiedijk-2008}:

\begin {quote}
``Most mathematicians are not aware that this third revolution already has happened, and many probably will disagree that this revolution even is needed.
However, in a few centuries mathematicians will look back at our time as the time of this revolution. In that future most mathematicians will not
consider mathematics to be definitive unless it has been fully formalized.''
\end {quote}

John Harrison also notes, in the conclusion of \cite {harrison-2008}, that:

\begin {quote}
``The use of formal proofs in mathematics is a natural continuation of existing trends towards greater rigor.''
\end {quote}

Proof assistants enable, first of all, the user to describe the objects of his/her universe of discourse. Then, to describe their behaviour and, finally, to prove properties exhibited by these objects in the form of mathematical lemmas and theorems. All this, however, must be done in a proper mathematical language, with a friendly syntax and a consistent and precise semantics. 

In the case of the Coq proof assistant, this theory is called ``Calculus of Inductive Constructions''. It is a type theory that extends the Curry-Howard Correspondence between terms of the typed lambda calculus and proofs of natural deduction, as well as between specifications (type expressions) and propositions. All these topics are briefly reviewed below.

\section {Logic and Natural Deduction}
Considered by some the foundations of mathematics, and by others as a branch of mathematics itself, logic is concerned, in general, with the representation and validation of arguments. For this purpose, formal logic comes in different flavors, such as propositional logic (whose sentences consist of variables grouped around a small set of connectives) and first-order logic (whose sentences allow the use of the universal and the existential quantifiers) \cite {dalen-2008}. Since mathematical theorems are generally formulated as propositions, it is a natural choice to use the language of first-order logic to express them formally.

``Classical logic'' refers to a logic where the law of the excluded middle is accepted. In this kind of logic, every proposition is either true or false, even if there is no proof of one or the other. ``Intuitionistic logic'', on the other hand, was originally conceived by Brouwer and further developed by Heyting and Gentzen, and restricts classical logic by not allowing the use of the law of the excluded middle. In this logic, for a proposition to be considered true or false there must exist a proof or one or the other. As an example, in classical logic the proposition from complexity theory ``P=NP or P different NP'' is valid, although it is not known, for the time being, whether either of the disjuncts is true or false. Intuitionistic logic, on the contrary, does not consider this as a valid proposition, as no one yet knows of a proof of either one of the disjuncts. 

Although not all propositions that can be proved in classical logic can also be proved in intuitionistic logic, the latter has the special property that from a proof one can effectively build the object that confirms the validity of the proposition (the proof behaves like a kind of a recipe for building such object). For this reason, intuitionistic logic is also called ``constructive'' logic. An important consequence of this is that computer programs can be extracted from constructive proofs, and for this reason this logic is known for having ``computational content'', a property that, although being extensively investigated via continuations or negative translations, does not naturally arise out of classical logic. Thus, it is of great interest in computer science.

In the present context, propositions will be used to specify computer programs (for example, properties of the computation performed by a program, relations between input and output data of a program etc) and also to state theorems and hypotheses. Both activities are indeed equivalent, as prescribed by the Curry-Howard Correspondence presented next. Nevertheless, programs must be built and theorems must be proved, and one important tool for this comes from proof theory and is known as ``Natural Deduction''.

Natural deduction uses a small set of inference rules in order to find a logical path from hypotheses (or none) to a conclusion that needs to be proved. The reasoning steps used in this process (directly related to the structure of the conclusion) are very simple and naturally related to common reasoning. The proof, in this case, is a tree that shows all the primitive reasoning steps used (one for each different connective) and how they transform the hypothesis into the conclusion.

\section {Untyped Lamba Calculus}
The untyped lambda calculus is a formal system used in the representation of computations \cite {hindley-2008}. It is, thus, a model of computation. It is based on the abstraction and application of functions, which are considered as higher order objects. This means that functions can both be passed as arguments to other functions, and can also be returned as values from function calls.

The so-called lambda language is a notation for the representation of lambda terms, which are then manipulated by the rules of the calculus. The power of the calculus is related to the simplicity of the lambda language, which uses only two operators to represent all computations. These are the function abstraction and the function application (call) operations. 

The Lambda Calculus was invented by the American mathematician Alonzo Church in the 1930s, as a result of his research on the foundations of mathematics. His objective, by then, was to formalize mathematics through the use of ``functions as rules'' instead of the classical use of sets as in set theory. Although he did not fulfill his objective, his work was of high importance to computer science, specially in the theory, specification and implementation of computer programming languages (the functional ones have a direct influence of it), representation of computable functions, computer program verification and proof theory. The first undecidable mathematical problems ever proved to be so in the scientific literature were formulated via untyped lambda calculus, even before the invention of Turing machines, which became since then a preferred model for computability research.

The rules of computation in the lambda calculus are also very simple, and can be grouped in a few cases. The idea of substitutions leads to the concept of reductions, which allows for the transformation of terms into equivalent ones. These reductions represent computations, and the equality of terms represents the equality of functions and values as we know from classical mathematics. Thus, a lambda expression represents a program, in the sense that it embeds an algorithm, takes arguments as inputs and produces a result. A reduction, on the other hand, represents a computation step, a change in the state of a computation. The result of computation is represented by a term that cannot be further reduced, and for this reason is called a ``normal form''.

This simple language and theory is powerful enough to represent data types as we know from programming languages, such as booleans and natural numbers, and also to represent the operations usually applied to them. Thus, lambda expressions can be seen as abstract representations of programs, with the same computational power one would expect from them.

Despite the simplicity and elegance of the untyped lambda calculus, it was proved that equality of lambda terms cannot be decided algorithmically. Also, it cannot be decided if an untyped lambda term has a normal form or not. These results, discovered by Church, were the first undecidable problems to be proved in  history. They were also used to prove the undecidability of first order pure predicate logic.

\section {Typed Lambda Calculus}
The typed lambda calculus is a variant from the untyped version, with type tags associated to the lambda terms \cite {hindley-2008}. Thus, all atomic elements have a type, and the type of a lambda term can be inferred from the type of its subterms and atomic elements. Types, in this case, are used to ensure that only arguments of the correct type are passed to functions, and that functions are used in a consistent way inside other functions. The intention is to avoid inconsistencies and paradoxes, as they might occur in the untyped lambda calculus.

As a consequence, the typed lambda calculus is a less powerful model of computation, but offers type checking capabilities that can be used to ensure the correct use of applications and abstractions. When applied to computer programming language development, this leads to safer languages that can detect more errors, either statically or dynamically. Indeed, most of the safe languages used nowadays for computer programming take advantage of this type feature and implement it in different ways.

Differently from the untyped lambda calculus, equality of lambda terms is decidable in the typed version. Also, it is decidable if a term has a normal form or not. It can be checked, for example, if two programs are equal (in the sense that they produce the same results for the same inputs) and also if a program produces a result at all for a given input (in this case it is enough to check if the correspondent expression reduces to a normal form). For these reasons, there is a lot of interest in the use of the typed lambda calculus as a model of computation, despite this model being more restricted than the one obtained with the untyped version.

In what follows, ``computer programs'' is a synonym for ``lambda terms'' (typed or untyped, depending on the context), and ``computation'' means the application of a series of reductions to a lambda term (idem).

\section {Curry-Howard Correspondence}
There is no reason, in principle, to consider that the two fundamental aspects of mathematics - reasoning and computing - are related in any way.  This is not the case, however, as stated by the Curry-Howard Correspondence (also known as ``Equivalence'' or ``Isomorphism'') \cite {sorensen-2006}.

Initially observed by Haskell Curry, and later developed by William Howard, the correspondence establishes a direct relationship between logic and computation models (the typed lambda calculus in the present context).

On one hand, it is a fact that every typed lambda term can be associated to a type expression, which expresses its type. This type expression, in our context, is called a ``specification'', because it denotes the signature of the correspondent program (i. e., the number and types of the arguments and the type of the result). On the other hand, it is also a fact that the proof of a proposition, represented by the tree built by a natural deduction process, puts one (the proof) in direct relation to the other (the proposition). Thus, (typed lambda) terms have specifications (types), and proofs (natural deduction trees) confirm the validity of propositions (theorems).

The Curry-Howard Correspondence is a syntactic analogy between terms and proofs, and also between specifications and propositions, which has deep consequences in the way one can reason about programs and develop proofs of theorems. Basically, it states that proofs and terms have a direct relationship: a term that satisfies a specification (that is, a typed lambda term that has a certain type) is, at the same time, a representation of a proof in natural deduction of the proposition represented by the specification. The converse is also true: the proof of a proposition can be taken as a term whose specification is the proposition itself.

Going further, the Curry-Howard Correspondence establishes that to prove a theorem (proposition) is the same as finding a term whose type (specification) matches the proposition. Thus, finding a proof of a theorem is the same as building a program. On the other way around, to build a term that satisfies a certain specification is the same as proving the proposition that matches this specification. According to it, to build a proof of a proposition is the same as to build a program of a specific type, where the proposition represents the type of the program. 

Another important consequence is that proof validation can be done via simple program type checking routines, which is usually an easy and efficient task. In summary, the Curry-Howard Correspondence states that building a program is the same as finding a proof, and both activities are interchangeable and highly interconnected. 

The Correspondence opens up an immense area of research by exploring the relationships between programs and proofs and also by specifications and propositions. In particular, it allows program development to be considered as a formal mathematical activity, with all the related benefits. Also, it allows for computer programming skills to be introduced in the process of theorem proving, with many advantages as well.

\section {Type Theory}
Type theories are formal systems that can serve to represent the foundations of mathematics. They make extensive use of the idea of ``types'', which are abstract tags that are assigned to values, specially to the values passed as arguments to functions and to the values returned by function calls. This way, type checking can be performed on the terms constructed in these theories, and this helps avoiding some of the inconsistencies found in other base theories, such as set theory. The typed lambda calculus of Church is, itself, a type theory, one of the first ones to be formulated.

Type theory has its origins before Church, however. It was invented by Bertrand Russell in the 1910s, as a tentative of fixing the inconsistencies in set theory, as a consequence of Russell's Paradox stated in the beginning of the 20th century (``is the set composed of all sets that are not members of themselves a member of itself?'')

Since then, many different type theories have been created, used and developed. Martin-L\"of's Intuitionistic Type Theory \cite {martin-1980} \cite {nordstrom-1990} is one of the most important, and is widely used as the basis of modern constructive mathematics and also as the fundamental theory of many proof assistants. Derived from the mathematical constructivism that appeared in the end of the 19th century, the main characteristic of the constructivist ideology is, according to Troelstra \cite {troelstra-2011}:

\begin {quote}
... the insistence that mathematical objects are to be constructed (mental constructions) or computed; thus theorems asserting the existence of certain objects should by their proofs give us the means of constructing objects whose existence is being asserted.
\end {quote}

Martin-L\"of's Intuitionistic Type Theory is based on the first-order logic to represent logical propositions and types, and the typed lambda calculus to represent programs and proofs of theorems. The whole idea is structured around the Curry-Howard Correspondence, which relates proofs to programs and propositions to types.

In a constructive type theory \cite {thompson-1999}, propositions are not taken for the truth value that they represent - either true or false, as in the Tarski tradition. Instead, the BHK (Brouwer, Heyting and Kolgomorov) interpretation is used, which asks for the proof of the validity of a statement. In it, every true proposition must be accompanied by a corresponding proof, and every logical connective used to build a new proposition is associated to an inference rule that is used to build the proof of the new proposition. A proof of $A \land B$, for example, is a pair $<a,b>$ where $a$ is a proof of $A$ and $b$ is a proof of $B$. In the same way, every logical connective has an associated inference rule. Different rules apply for other logical connectives such as disjunction, implication and negation, and also the universal and existential quantifiers.

\section {Calculus of Constructions with Inductive Definitions}
The Calculus of Constructions with Inductive Definitions (previously known as Calculus of Inductive Constructions) \cite {coq-2012} is an extension of the Calculus of Constructions developed by Thierry Coquand \cite {coquand-1986}, and is a constructive type theory that was designed specifically to be used as the mathematical language of the Coq proof assistant. 

With it one can use logical connectives and define mathematical objects and their corresponding types (naturals, integers, chars, strings etc), and then reason about them with inference rules and functions. After some time, an induction mechanism was added to it, allowing types and propositions to be defined inductively, along with mechanisms to reason about these inductively defined objects, thus leading to the name Calculus of Inductive Constructions. Basically, it is a typed lambda calculus with a rich type system extended with inductive features.

\chapter {Proof Assistants}
\label {cha-pa}
Proof assistants (also known as ``interactive theorem provers'') are computer-based software tools that help users to state and proof results in a certain theory \cite {geuvers-2009}. They date from as far as 1967, when De Bruijn launched the Automath project \cite {bruijn-1983}, and their development follows closely the development of computers and computer software themselves. Since then, they have evolved a lot and many such systems are available today \cite {wiedijk-2006}. Among the most known and influential, it is possible to mention Coq \cite {bertot-2004} \cite {coq-2012} \cite {coq-site}, Agda, HOL, Isabelle, Matita, Mizar and Nuprl. Each has its own specific features, is based on a particular mathematical calculus and some of them have inspired the development of the others. A brief history of their main characteristics, usage, evolution and perspectives can be found in \cite {geuvers-2009}.

Basically, a proof assistant is an interactive tool that allows the user to write expressions that represent propositions/types and proofs/terms in its input language. Proofs/terms can then be interactively built from propositions/types, and/or check against them. Proof/terms can be built directly or indirectly, in the latter case via a command language (also called ``tactics''). They are then analyzed and checked for errors. Feedback is given via error messages, if an error is found. Otherwise, the proofs/terms are partially built by the result of applying inference rules to the current goal in a proof session. New subgoals might be generated as the session continues. When no more subgoals are generated, the proof/term is completed. Proved theorems can then be saved and used later in other contexts. Libraries containing a whole theory can be built, compiled and deployed elsewhere. Some proof assistants offer a graphical user interface, with windows, menus and so on, in order to increase user productivity.

In certain situations, a formal proof might involve some level of ``automated theorem proving'' (when the computer tries to prove at least part of the theorem without any help), but this is not usual since the process is normally guided by the strategy and hints provided by the user. On the other hand, in all cases it involves ``automated proof checking'', which is a much simpler process of checking that a proof is correctly constructed according to the inference rules of the underlying theory.

The term ``assistant'' stresses the fact that such tools area not unrealistic automated theorem provers, but instead assistants for a human that needs to state and prove a theory with a number of lemmas and theorems, checking all steps of the development and offering some high level actions that ease the construction of the proofs. This is done in a setting of an underlying mathematical theory, whose characteristics drive the syntax and semantics of the language used and also shapes the style of reasoning and computing that can be used to achieve the results. For this reason, one must first consider the nature, the roots, the limits and the capabilities of this theory before considering its use in a computational environment.

``Proof irrelevance'' is an important concept when it comes to using a proof assistant to automate, as much as possible, the construction of proofs and terms. Since the importance of a proof relies simply on its existence, no matter how long or complex it is, the use of a proof assistant should place no burden when some automation is done in the process. This means, essentially, that there are no important differences between two different proofs of the same proposition. On the other hand, finding a term that complies with a specification might have serious impact on its usage as a computer program, since readability and efficiency are standard concerns. Two different terms might verify the same specification, but for practical reasons there might be important differences between them. For this reason, proof assistants are not generally considered when pursuing automated program development.

Proof assistants help the user to formally prove his/her theorems and/or build his/her programs. But what is a formal proof anyway? As noted by Thomas Hales in \cite {hales-2008}, a formal proof is not only a proof that is written in a precise and non-ambiguous mathematical language, but also a construction that can be verified, in every step, according to the fundamental axioms of mathematics. As this involves a lot of effort, not usually fully undertaken by the practicing mathematician in the so-called informal proofs, the computer and the proof assistant play an important role in making this a feasible task. In his own words:

\begin {quote}
``A formal proof is a proof in which every logical inference has been checked all the way back to the fundamental axioms of mathematics. All the intermediate logical steps are supplied, without exception. No appeal is made to intuition, even if the translation from intuition to logic is routine. Thus, a formal proof is less intuitive, and yet less susceptible to logical errors.''
\end {quote}

The main benefits of using a proof assistant to develop mathematical theories and computer software are:
\begin {itemize}
\item A powerful and uniform language is provided along the whole process of representation and proofing;
\item Programs can be formally checked against their specifications;
\item Program verification can be reduced to a simple type checking algorithm that can be very fast and reliable;
\item Reduced user time and effort necessary to check proofs and programs, specially long and complex ones, as these are completely carried by the machine;
\item Because of the above, the checking process is also more reliable and less error-prone;
\item Equivalent proofs can be considered quantitatively, that is, against criteria such as readability, level of abstraction, size, structure etc;
\item Results can be stored for future reuse and can be easily retrieved in different contexts;
\item Different approaches can be easily tested, and results easily combined;
\item A proof assistant can interact with the user in order to ease the construction of the proof or the program, sometimes doing all or part of the work;
\item The user gets a deeper insight into the nature of the proof or program, enabling a better understanding of the problem and usually leading to simpler or more efficient solutions.
\end {itemize}

As pointed out by Georges Gonthier in \cite {gonthier-2008b}, an important benefit of developing a formal proof is not only to become sure of its correctness. Due to the level and quantity of details involved, it is also an important exercise in getting a better understanding of the nature of the proof, possibly leading to further simplification.

Current applications of proof assistants include the formalization of important mathematical theorems, but are not limited to these. Relevant and more pragmatic uses include also the automatic review of large and complex proofs in papers submitted to specialized journals, and the verification of hardware and software components in computer systems, as noted by John Harrison in \cite {harrison-2008}. The paper also brings a concise presentation of the foundations, history, development and perspectives of the use of interactive proof assistants.

A general claim against the use of proof assistants is the unreliability of the results, caused by possible errors in the underlying infrastructure. This includes bugs or failures in the code of the proof assistant itself, the processor of the high level language used to implement it, as well as related libraries, the operating system and the hardware on which it is executed. This, however, can be virtually eliminated as these infrastructure items are not specific to proof assistants. On the contrary, they are usually commercial and large-scale products that are extensively used and tested in wide variety of situations and long periods of time, thus making it very unlikely to confirm the worries of the claim.

Another important aspect of formal proofs is that their size can be much larger than the corresponding size of the original informal proofs. This is due to the very detailed nature of these proofs and, as discussed by Freek Wiedjik in \cite {wiedijk-2000}, this expansion factor seems to be the constant four. He came to this conclusion after comparing the size of the formal and informal proofs of different important projects in different proof assistants. The factor became since then known as the ``de Bruijn factor'' in honor to Nicolaas Govert de Bruijn who first observed this phenomenon in the early 1980s while working in the Automath project.

Proof assistants are normally implemented in some functional language (like Lisp, Haskell, Ocaml and ML) and differ in some fundamental features, such as the support for higher-order logic, dependent types, code extraction, proof automation and proof by reflection. Most of them are developed and maintained by research institutes and universities in Europe and USA, and can be freely used and distributed. 

Freek Wiedijk compiled in 2006 a list of what he calls ``The Seventeen Provers of the World'', with the objective of gathering information on and comparing the features of the main provers available by then \cite {wiedijk-2006}. For this, he invited researchers with different formal proof backgrounds and expertise to provide a formal proof of $\sqrt 2$ being irrational in different proof assistants, in order to highlight their usage style, expressive power, limitations and capabilities. Also, the paper includes a profile of each of these assistants and brings much insight into their nature and objectives, as well as their differences and unique characteristics.

Proof assistants might lead, in the future, to the fulfilment of the dream described in ``The QED Manifesto'' \cite {qed-1994}. Written and published by an anonymous group of researchers dedicated to automated reasoning in 1994, the text claims for a single automated repository of all mathematical knowledge and techniques, where the contents are previously mechanically checked and then made freely available to researchers and students all over the world. Although important from an historical perspective, the practical consequences of this initiative were very limited: only a mailing list and two conferences (1994 and 1995) were organized following its publication. The reasons for this were analyzed by Freek Wiedijk 13 years later in \cite {wiedijk-2007}, and among the most relevant he lists (i) the reduced number of people working with formal mathematics by then and (ii) the fact that, according to his opinion, ``formal mathematics does not resemble mathematics at all'', since ``formal proofs look like computer program source code'', unsuitable for use by pure mathematicians.

Surely much still has to be done, but the recent developments in the area of formal mathematics, along with a wider use of related tools and specially the increasingly large and complex projects that were undertaken since then, all point to the relevant role that the area will play both in mathematics and computer science over the next years. The QED Manifesto dream might not be too far in the future after all.

\chapter {The Coq Proof Assistant}
\label {cha-coq}
The Coq proof assistant was initially developed at INRIA by G. Huet and T. Coquand in 1984, and the first version was released in 1989. In 1991, with the addition of inductive types, it was renamed Coq. Since then, many other people have contributed to its development, adding increasingly new features and functionalities to the tool. The whole development history is described in the Coq Reference Manual \cite {coq-2012}. In 2013, Coq received the Programming Languages Software Awards from the ACM for its ``significant impact in programming language research, implementations and tools'' \cite {coq-award}.

Coq was designed to allow the representation of theorems and specifications (type expressions), and also to develop proofs of these theorems and the construction of certified programs (programs that satisfy their specification). Proofs and programs can be constructed directly, using a language called \emph {Gallina} (which is based on the Calculus of Inductive Constructions), or can be developed interactively using a set of ``tactics'' that assist the user in this process. Gallina is a dependently typed functional programming language that combines a higher-order logic language and a typed functional language, and is used to represent terms/proofs and types/propositions in the system. Another language, called \emph {Vernacular}, is used to interactively assist the user in the development of his/her proofs. 

Tactics are commands that, when interpreted, apply one or more rules of the Calculus of Inductive Constructions in the current context to the current goal. The initial goal is the proposition to be proved, which can be a single theorem (or lemma) of the specification of some property of a program. The application is in reverse order, so that the current goal is substituted by one or more (simpler) subgoals and possibly some changes to the context. When no new subgoals are generated, the initial goal is considered valid and a proof term, constructed from the application of the rules according to the tactics used, is built.

A type-checking algorithm ensures that proofs are correctly constructed and, consequently, that programs satisfy their specifications. In this sense, Coq is suited not only for developing abstract mathematical theories, but also for the development of correct programs and the certification of already developed programs. As discussed in the next chapter, many important theories and systems have been successfully developed over the recent years with Coq.

Coq uses a constructive logic and comes with a large standard library that defines fundamental data types (natural numbers, integers, booleans, lists etc), implements most operations on them and proves their most important properties \cite {coq-lib}. Also, there is a large set of user contributions from the most diverse areas, which can be freely accessed and used \cite {coq-contrib}. Finally, Coq can be extended with different plug-ins that offer the user alternative (usually higher-level) script languages (e.g. SSReflect \cite {whiteside-2012}).

Theories can be compiled and stored in object files for efficient reuse. Also, Coq supports program extraction from terms/proofs into different commercial computer programming languages.

Coq is implemented in the Ocaml functional programming language and supports higher-order logic, dependent types, code extraction, proof automation and proof by reflection, thus being a powerful tool suited for a wide range of applications. Despite its many advantages, some critics consider the foundations of Coq too complicated to be understood and be used by mathematicians. Also, the facts that Coq's logic is constructive (except in a few cases, one cannot use the Law of the Excluded Middle) as opposed to classical logic, and also that equality is intensional, as opposed to extensional (where different reduction strategies do not have to be considered), are considered unrealistic and too distant from the universe of the practicing mathematician \cite {wiedijk-2007}. Nevertheless, Coq is nowadays considered one of the most important interactive theorem provers around and is being deployed in significant formalization projects, as discussed in the next chapter.

\chapter {A Sampler of Formally Checked Projects}
\label {cha-sampler}
As a consequence of the increasing interest in computer-based formal mathematics over the recent years, new and relevant uses of Coq are announced frequently. These include a wide range of projects in programming language semantics formalization \cite {coq-projects-lang}, mathematics formalization \cite {coq-projects-math} and teaching in both mathematics and computer science curricula \cite {coq-projects-teach}. Most projects are academic, but there are also important industrial applications. Some of these (three more theoretical and three more application-oriented, five in Coq and one in Isabelle) are presented in this chapter. These should convince the reader of the rapidly increasing importance of doing formal mathematics and using interactive theorem provers in current and future practice.

Besides that, Freek Wiedijk keeps a list that shows the formalization status of 100 important mathematical theorems and includes, for example, Fermat's last theorem. The list is updated continuously and is based on another list with the same theorems, chosen for ``the place the theorem holds in the literature, the quality of the proof, and the unexpectedness of the result'' \cite {wiedijk-top100}. In 2008, according to Wiedijk in \cite {wiedijk-2008}, 80\% of the theorems in this list had been formalized in at least one proof assistant, and he expressed his belief that this number would reach 99\% in the next few years. This however has not yet happened, as the percentage informed by the site in January of 2014 is only 88\%. The most used proof assistants in this list are, in decreasing order, HOL Light, Mizar, Isabelle and Coq. Some simple theorems, such as the irrationality of $\sqrt {2}$, have more than a dozen different proof formalizations, developed in different proof assistants.

\section {Four Color Theorem}
It took one hundred and twenty four years since the Four Color Theorem was stated until it was finally was proved (1852-1976). It is an important theorem because (i) it has a simple statement:
\begin {quote}
The regions of any simple planar map can be coloured with only four colours, in such a way that any two adjacent regions have different colours.
\end {quote}
and (ii) it was the first famous mathematical theorem to be proved with the aid of computers. These proofs date from 1976 and 1995, when different mathematicians provided, respectively, an initial proof that was very large and complex, and a more concise and readable demonstration of it. Both relied on the use of computers to analyze hundreds of cases in the demonstrations. Because of this new way of proving mathematical theorems, the 1976 proof was very much criticized for those that considered computers unsuited for this task for a number of reasons, ranging from ``unnatural'' to ``unreliable''.

Nevertheless, the 1995 proof still relied on partial human intervention, and for this reason cannot be considered a fully computer-checked proof. This happened only ten years later, in 2005, when Georges Gonthier, then a principal researcher from Microsoft Research, with the collaboration of Benjamin Werner of INRIA (\emph {Institut National de Recherche en Informatique et en Automatique}, the French research institution fully dedicated to computational sciences), used the Coq proof assistant to develop a proof script that was completely verified by a machine \cite {microsoft-four} \cite {gonthier-2008} \cite {gonthier-2008b} \cite {gonthier-four}.

Considered a milestone in the history of computer assisted proofing, the importance of this demonstration can be summarized in Gonthier's own words \cite {gonthier-four}:
\begin {quote}
``Although this work is purportedly about using computer programming to help doing mathematics, we expect that most of its fallout will be in the reverse direction — using mathematics to help programming computers.''
\end {quote}
In the same reference, he expresses the fundamental role that is reserved for mathematics as a software development tool, with the aid of computers, and exposes new and exciting possibilities and directions for formal mathematics and computer technology evolution.

With approximately 60,000 lines of code and 2,500 lemmas in the proof script, Gonthier points that \cite {gonthier-four}:
\begin {quote}
``As with most formal developments of classical mathematical results, the most interesting aspect of our work is not the result we achieved, but how we achieved it.''
\end {quote}

The effort in this project also led to a byproduct, an extension to the Coq scripting language in the form of a new set of tactics designed to ease and reduce the development effort by implementing a style of proofing known as ``small scale reflection''. This set turned out to be a plugin for the Coq system, which is now available under the name SSReflect \cite {whiteside-2012}.

An important example of the cooperation between Microsoft Research and INRIA (among others), this work was finalized in the same period when these organizations decided to share efforts and create a new initiative dedicated to research on projects of common interest. This culminated in an association between them that was formalized with the inauguration of the Microsoft Research-INRIA Joint Centre in January of 2007 \cite {microsoft-inria}. Almost a year before it was inaugurated, three formal methods projects were launched and, since then, a wide selection of projects is supported, mainly in the areas of the application of mathematics to increase the security and reliability of software and computer systems, and also on the development of new software tools and applications for complex scientific data. Georges Gonthier is nowadays a researcher of the Joint Centre.

\section {Odd Order Theorem}
Right after finishing the proof of the Four Color Theorem, Georges Gonthier engaged in an even more ambitious project: the formal proof of the Odd Order Theorem, also known as the Feit-Thompson Theorem. Conjectured by Burnside in 1911, it was first proved by Walter Feit and John G. Thompson in 1963. The work of Gonthier, of enormous importance, started in May of 2006 and ended in September of 2012, thus consuming more than six years of a team led by him with full time dedication. Besides Gonthier, a number of other people from Microsoft Research and INRIA in several different locations worked together in order to achieve the result that rapidly gained attention throughout the world \cite {microsoft-inria-feit} \cite {microsoft-inria-feit2}.

The theorem has deep implications both in mathematics (in the classification of finite simple groups) and also in computer science (specially in cryptography). Although the theorem states shortly (``any finite group of odd order is solvable''), its formal proof is approximately 150,000 lines long and contains some 13,000 theorems (almost three times the size of the Four Color theorem formal script). These big numbers are not due to the formal approach used, since the original demonstration by Feit-Thompson itself was quite long, with 255 pages of text.

The announcement by Microsoft Research (\cite {microsoft-feit}) brings interesting opinions by Gonthier, stressing the context, the importance and the consequences of his work :

\begin {quote}
``The work is about developing the use of computers as tools for doing mathematics for processing mathematical knowledge. Computers have been gaining in math, but mostly to solve ancillary problems such as type setting, or carrying out numerical or symbolic calculation, or enumerating various categories of common natural objects, like polyhedra of various shapes. They're not used to process the actual mathematical knowledge: the theories, the definitions, the theorems, and the proofs of those theorems. My work has been to try to break this barrier and make computers into effective tools. That is mostly about finding ways of expressing mathematical knowledge so that it can be processed by software.''
\end {quote}

In the same announcement Andrew Blake, director of Microsoft Research Cambridge, summarizes:

\begin {quote}
``One may anticipate that this could affect profoundly both computer science and mathematics.''
\end {quote}

The whole development is presented in detail in \cite {gonthier-feit} \cite {gonthier-feit2}. As pointed out in the conclusion by Georges Gonthier,

\begin {quote}
``The outcome is not only a proof of the Odd Order Theorem, but also, more importantly, a substantial library of mathematical components, and a tried and tested methodology that will support future formalization efforts.''
\end {quote}

\section {Homotopy Type Theory and Univalent Foundations of Mathematics}
The Univalent Foundations Program is an ambitious research program developed at the Institute for Advanced Study in Princeton by group led by Vladimir Voevodsky. It aims at building new foundations for mathematics, as an alternative to classical set theory, and is based on a new interpretation of Martin Löf's constructive type theory into the traditional field of homotopy \cite {awodey-2012} \cite {pelayo-2012}. 

Lots of efforts are being put into this promising project, which now has the collaboration of many researchers from different parts of the world. Large part of the development is done in a modified version of Coq (another proof assistant, Agda, is also used) and, contrary to usual practice, many theorems are first formally proved and then converted to an informal description \cite {hott-2013}. Although considered a work in progress, four important sources compile the main results obtained so far: the official Homotopy Type Theory site \cite {hott-site}, the GitHub repository where the development is done \cite {github-site}, the arXiv page \cite {arxiv-site} and the collaborative book ``Homotopy Type Theory: Univalent Foundations Program'', freely available for download \cite {hott-2013}.

\section {Compiler Certification}
Moving from strictly academic or abstract projects into more realistic and pragmatic ones, at least in the short term, it is possible to find some important commercial initiatives, such as the CompCert project led by Xavier Leroy \cite {leroy-2009} \cite {compcert}. CompCert is a shorthand for the words ``Compiler'' and ``Certification'', and refers to a compiler that was fully verified by a machine in order to assure its semantic preservation properties: in other words, that the low-level code produced as output has, in every case, the same semantics as the high-level code taken as input.

Developed to be used in the programming of critical software systems, for which exhaustive testing is a very difficult task, CompCert brings new perspectives into the construction of safety-critical software for embedded applications, while generating code that still meets the performance and size criteria required for this type of usage. To be certified, in this case, means that all internal phases of the compiler were proved to implement the correct translations, including complex ones such as code optimizations. Thus, assuring that source and object code behave in exactly the same way can be extremely useful when, for example, formally proving that the source code satisfies all the required specifications, because these will also be satisfied by the executable code. Otherwise, a formally certified source program compiled by a non certified compiler could, due to programming errors in the implementation of the compiler, lead to code that did not comply with the specifications.

CompCert translates a large subset of the C programming language into the PowerPC machine code, the choice of both being related to their popularity in real-time applications, specially avionics. CompCert was not only certified in Coq, but it was fully developed (programmed) in the proof assistant as well, thus making the whole project self-contained and easier to reason about. With the help of Coq's code extraction facility, functions and proofs could easily be transformed into functional programs, thus leading to an efficient implementation. The whole project consumed three persons-years over a five years period, and comprises 42,000 lines of Coq code.

\section {Microkernel Certification}
Generating certified low level code is important, but running this code on a certified environment is of no less importance in order to assure that the expected behaviour is achieved.

Operating systems are critical components in every computer system and must be trusted in all situations, as they provide an execution environment --- supposedly reliable --- for application programs. Among its building blocks are microkernels programs that execute on the privileged mode of the hardware, meaning that detecting and recovering from software faults in this case is extremely difficult. This, of course, explains the interest in their mechanized certification, which in this case means making sure that the final implementation conforms exactly to the properties of the corresponding high level specifications.

The seL4 project was developed with this objective \cite {klein-2010}. Consisting of approximately 10,000 lines of C code, it was formally certified with the Isabelle/HOL proof assistant and, as a result, it claims to exhibit ideal characteristics for this kind of program: no crash and no execution of any unsafe operation (code injection, buffer overflow, general exceptions, memory leaks etc), under any circumstances. The cost of the certification was not low, indeed: some 200,000 lines of Isabelle script had to be used or developed in order to complete the proof, and consumed around 11 persons-years. The authors believe, however, that with the experience and lessons learned from this project, the size of new and similar projects could be reduced to 8 persons-years, which is the double of the estimate for equivalent projects without any kind of certification. This means, in practice, that formal certification is indeed a real possibility, even for a commercial product with thousands of lines of code, considering the benefits that can be obtained. It is also interesting to note, in their work, that they considered not only the cost of a full verification, but also the incremental cost of repeating the verification when some feature is added or changed to the microkernel, which is a realistic attitude for projects that evolve over time.

\section {Digital Security Certification}
Even before the Four Colors Theorem was mechanically proved, an important project of digital security verification was being conducted with the support of the Coq proof assistant.

Java Card$^{TM}$ is one of the most important smart card platforms being used. It allows users to carry with them, in a supposedly safe way, different applications that keep and share personal data, such as banking, credit card, health and many others. Applications run on a Java Card platform, and they can be preloaded or even loaded after the card has been issued to the user. 

Although the convenience and relevance of this technology is unquestionable, it raises problems that must be adequately dealt with, specially related to the confidence and integrity of the data stored in the card. In a context of multiple applications distributed by different companies over different periods of time, ``confidence'' means that sensitive data cannot be shared without authorization, and ``integrity'' means that data cannot be changed without permission. Achieving high levels of security in such an environment is not an easy task, but is certainly a must in order to enable its widespread and safe use, without exposing one's personal data among other possible harmful consequences.

For this reason, it is quite natural that efforts have been directed at formalizing the behaviour and properties of the Java Card platform, including the virtual machine, the runtime environment, the API and the Java Card programs themselves, in interactive theorem provers such as Coq, where a full machine-checked certification could be pursued.

This happened, indeed, with the work of many and different researchers and private companies in the beginning of the 2000s. Barthe, Dufay, Jakubiec and Sousa at INRIA, for example, published a series of papers that address the problem from different perspectives: from the formalization of the Java Card Virtual Machine \cite {barthe-2000} and other components \cite {barthe-2001} to the construction of offensive (more efficient) virtual machines from defensive (less efficient) ones \cite {barthe-2002}.

The industrial segment (banks, insurance and credit card companies) has invested a lot in this technology and also put direct efforts in the verification of  its safety properties. An example of this is the paper by Andronick, Chetali and Ly from the Advanced Research Group on Smart Cards at Schlumberger Systems, which used Coq to prove applet isolation properties in the Java Card platform \cite {andronick-2003}. Applet isolation, in this case, means that neither the platform will become damaged, nor an application will interfere with another in an improper way. Another initiative was published in 2007, when Gemalto, an important developer and supplier of digital security technology for financial institutions, telecom operators and governments all over the world, announced that it had achieved, for the first time in history, the highest possible level of certification for its Java Card implementation, after extensive formalization with the Coq proof assistant \cite {javacard}.

\chapter {Context-Free Language Theory}
\label {cha-lang}
Language and automata theory is a long standing and well developed area of knowledge that uses mathematical objects and notations to represent devices that constitute the basis of computation and computer technology, both in software and hardware. It has its roots in the then independent automata theory and language theory, which became a single discipline in the 1950s, as a consequence of the research by Noam Chomsky and others. The classical theory is consolidated since the 1970s, and after that only specific research topics continue to evolve independently.

It comprises, initially, the representation and classification of phrase-structured languages, and then the study of its properties. In the Chomsky Hierarchy \cite {chomsky-1959}, the language classes considered are the regular, context-free, context-sensitive and recursively enumerable languages. For each of these classes, it is possible to study its representation alternatives (e.g. regular sets, regular expressions, linear grammars and finite automata for regular languages) and the equivalences between them. Also, it is possible the formulate and prove some properties, such as the existence of minimal finite automata, the decidability of some questions (such as whether a string is generated by a linear grammar) and the validity of some closure properties (e.g. the union of the regular languages is also a regular language). Other language classes not originally part of the Hierarchy, such as the recursive and the deterministic context-free languages, were also identified and can be the subject of the same type of study.

Language and automata theory is the basis of language representation and implementation in general, and has a direct impact in areas such as programming language specification and compiler construction. Other applications are in the area of data description, digital circuits, control systems and communication protocols design and natural language processing. Also, it is the basis of the study of fundamental areas of computer science, such as problem decidability and algorithm complexity. 

Although extensively studied in the past, with many classical books published since then \cite {hopcroft-1979} \cite {papadimitriou-1998} \cite {sipser-2005} \cite {sudkamp-2006}, these theories are still being explored in new areas of research, such as adaptive technologies, where extensive investigation, specially of its formal aspect, remain to be done. Nevertheless, new approaches to the representation and study of language and automata theory become available now and then. Examples are the development of JFLAP \cite {jflap-site} \cite {rodger-2006}, a graphical tool that allows the user to interactively experiment with abstract models, and other similar tools, graphical or not. Also, there are innovative contributions, such as the software engineering approach that models entities in terms of UML and Ruby language objects, in order obtain precise descriptions of its behaviour and enable to user to acquire a more practical view of the theories while being also able to experiment with it \cite {ramos-2009}.

Except for a few initiatives that have pursued a formal, mechanically checked, approach to language and automata theory (see Chapter \ref {cha-related}), most of the published literature in books and articles remain informal, in the sense that no uniform notation has been used, no mechanized verification of the proofs has been obtained and no certified algorithms have been constructed. This contrasts with the large interest in language and automata theory, which ranges from academy to industry, from students to researchers and practitioners. Thus, it seems natural that an effort of formalizing this theory should be beneficial for a wide public and the countless different applications. For this reason, the objective of this work is fundamentally to bring language and automata theory into the Coq proof assistant. Moreover, for the reasons that will become clear later, context-free language theory has been chosen as the starting point for this process.

The initial plan is to follow the structure of a book on formal languages and automata written by the author, together with João José Neto and Ítalo Santiago Vega \cite {ramos-2009}, and then to derive formalizations and machine checked proofs of its contents. Since the focus of the book is computational (proofs are substituted by algorithms), the idea of formalizing its contents seems attractive and a good complement for the book. Since the book adopts a software engineering approach, presenting UML models and Ruby implementations, merging these with Coq scripts could also give rise to further and fruitful research.

Formalizing language and automata theory can also be the starting point for the formalization of computability and complexity theory, areas of great interest in both academic and industrial contexts and where important research projects are being conducted.

\chapter {Current Status}
\label {cha-current}
Despite the great diversity and availability of literature and tools related to language and automata theory, it is not easy to find initiatives trying to describe this theory in a proof assistant, enabling the mechanization of theorem demonstration and derivation of certified programs that is becoming increasingly popular in all computer and mathematical areas.

The objective of this work is, thus, to mechanize language and automata theory into the Coq proof assistant, making it possible to reason about it in a fully checked environment. To achieve this goal, as with any other formalization, the following phases have to be accomplished:

\begin {enumerate}
\item \label {steps-select} Select an underlying formal logic to express the theory and then a tool that supports it adequately;
\item \label {steps-represent} Represent the objects of the universe of discourse in this logic;
\item \label {steps-implement} Implement a set of functions (or write a set of definitions) that provide basic transformations and mappings over these objects;
\item \label {steps-describe} Describe the properties and the behaviour of these objects and functions;
\item \label {steps-state} State lemmas and theorems that establish a consistent and complete theory;
\item \label {steps-derive} Formally derive proofs of these lemmas and theorems, leading to proof objects that can confirm their validity;
\item \label {steps-obtain} Obtain executable programs from these proof objects, such that they can be used for practical and efficient computation with the objects defined.
\end {enumerate}

Although the present report refers to an ongoing work, some interesting and important results have already been achieved. For phase \ref {steps-select}, the idea is to use the Calculus of Constructions with Inductive Definitions and the Coq proof assistant. Besides Coq having good support and a large community of users, with many important projects of high complexity already developed in it, the calculus is very powerful and the language is very flexible, as mentioned in most papers reporting formalizations in Coq. The remaining phases will depend on the theory being formalized.

Context-free language theory comprises a large number of different results. The focus of this work is on the formalization of the following ones:

\begin {enumerate}
\item Closure properties for context-free grammars (namely union, concatenation and Kleene star closure);
\item Grammar simplification (namely elimination of inaccessible and useless symbols, unit and empty productions);
\item Chomsky normal form for context-free grammars;
\item Decidability of the membership problem for context-free languages;
\item Pumping Lemma for context-free languages. 
\end {enumerate}

From these, closure and simplification have already been completed. The results are presented in the following chapters. 

\chapter {Closure Properties}

The formalization of closure properties for context-free grammars, namely the correcteness of the grammar union, grammar concatenation and grammar closure (Kleen star) operations is discussed in this chapter.

\section {Basic definitions}
\label {sec-formal-lang}
Context-free grammars have been represented in Coq very closely to the usual algebraic definition $G=(V,\Sigma,P,S)$, where $\Sigma$ is the set of terminal symbols (used in the construction of the sentences of the language generated by the grammar), $N=V-\Sigma$ is the set of non-terminal symbols (representing different sentence abstractions), $P$ is the set of rules and $S \in N$ is the start symbol (also called initial or root symbol). Rules have the form $\alpha \rightarrow \beta$, with $\alpha \in N$ and $\beta \in V^*$. The following record representation has been used:

\begin{verbatim}
Record cfg: Type:= {
non_terminal: Type;
terminal: Type;
start_symbol: non_terminal;
sf:= list (non_terminal + terminal);
rules: non_terminal -> sf -> Prop
}.
\end{verbatim}

The definition above states that \texttt {cfg} is a new type and contains four components. The first is \texttt {non\_terminal}, which represents the set of the non-terminal symbols of the grammar, the second is \texttt {terminal}, representing the set of terminal symbols, the third is \texttt {start\_symbol} and the fourth is \texttt {rules}, that represent the rules of the grammar. Rules are propositions (represented in Coq by \texttt {Prop}) that take as arguments a non-terminal symbol and a (possibly empty) list of non-terminal and terminal symbols (corresponding, respectively, to the left and right-hand side of a rule). \texttt {sf} (sentential form) is a list of terminal and non-terminal symbols.

Another fundamental concept used in this formalization is the idea of \emph {derivation}: a grammar \texttt{g} \emph {derives} a string \texttt {s2} from a string \texttt {s1} if there exists a series of rules in \texttt {g} that, when applied to \texttt {s1}, eventually result in \texttt {s2}. An inductive predicate definition of this concept in Coq uses two constructors:

\begin{verbatim}
Inductive derives (g: cfg): sf g -> sf g -> Prop :=
| derives_refl: forall s: sf g,
                derives g s s
| derives_step: forall s1 s2 s3: sf g,
                forall left: non_terminal g,
                forall right: sf g,
                derives g s1 (s2 ++ inl left :: s3)%list ->
                rules g left right ->
                derives g s1 (s2 ++ right ++ s3)%list.

\end{verbatim}

The constructors of this definition (\texttt {derives\_refl} and \texttt {derives\_step}) are the axioms of our theory. Constructor \texttt {derives\_refl} asserts that every sentential form \texttt {s} can be derived from \texttt {s} itself. Constructor \texttt {derives\_step} states that if a sentential form that contains the left-hand side of a rule is derived by a grammar, then the grammar derives the sentential form with the left-hand side substituted by the right-hand side of the same rule. This case corresponds the application of a rule in a direct derivation step.

Finally, a grammar \emph {generates} a string if this string can be derived from its root symbol:

\begin{verbatim}
Definition generates (g: cfg) (s: sf g): Prop:=
derives g [inl (start_symbol g)] s.
\end{verbatim} 

With these definitions, it has been possible to prove lemmas and also to implement functions that operate on grammars, all of which were useful when proving the main theorems. 

\section {Formalization}
After context-free grammars and derivations were defined, the basic operations of concatenation, union and closure of context-free grammars were implemented in a rather straightforward way. These operations provide, as their name suggests, new context-free grammars that generate, respectively, the concatenation, the union and the closure of the language(s) generated by the input grammar(s). The scripts for these terms is presented in the next sections.

\subsection {Union}

\begin{verbatim}
Definition g_uni_t (g1 g2: cfg): Type:= 
(terminal g1 + terminal g2)%type.

Inductive g_uni_nt (g1 g2: cfg): Type :=
| Start_uni : g_uni_nt g1 g2
| Transf1_uni : non_terminal g1 -> g_uni_nt g1 g2
| Transf2_uni : non_terminal g2 -> g_uni_nt g1 g2.

Definition g_uni_sf_lift_left (g1 g2: cfg)
(c: non_terminal g1 + terminal g1): g_uni_nt g1 g2 + g_uni_t g1 g2:=
  match c with
  | inl nt => inl (Transf1_uni g1 g2 nt)
  | inr t  => inr (inl t)
  end.

Definition g_uni_sf_lift_right (g1 g2: cfg)
(c: non_terminal g2 + terminal g2): g_uni_nt g1 g2 + g_uni_t g1 g2:=
  match c with
  | inl nt => inl (Transf2_uni g1 g2 nt)
  | inr t  => inr (inr t)
  end.

Inductive g_uni_rules (g1 g2: cfg): g_uni_nt g1 g2 -> 
list (g_uni_nt g1 g2 + g_uni_t g1 g2) -> Prop :=
| Start1_uni: g_uni_rules g1 g2 (Start_uni g1 g2) 
              [inl (Transf1_uni g1 g2 (start_symbol g1))]
| Start2_uni: g_uni_rules g1 g2 (Start_uni g1 g2) 
              [inl (Transf2_uni g1 g2 (start_symbol g2))]
| Lift1_uni: forall nt s,
             rules g1 nt s ->
             g_uni_rules g1 g2 (Transf1_uni g1 g2 nt) 
             (map (g_uni_sf_lift_left g1 g2) s)
| Lift2_uni: forall nt s,
             rules g2 nt s ->
             g_uni_rules g1 g2 (Transf2_uni g1 g2 nt) 
             (map (g_uni_sf_lift_right g1 g2) s).

Definition g_uni (g1 g2: cfg): cfg := {|
non_terminal:= g_uni_nt g1 g2;
terminal:= g_uni_t g1 g2;
start_symbol:= Start_uni g1 g2;
rules:= g_uni_rules g1 g2
|}.
\end{verbatim}

The first definition above (\texttt {g\_uni\_t}) represents the type of the terminal symbols of the union grammar, created from the terminal symbols of the source grammars. Basically, it states that the terminals of both source grammars become terminals in the union grammar, by means of a disjoint union operation. For the non-terminal symbols, a more complex statement is required. First, the non-terminals of the source grammars are mapped to non-terminals of the union grammar. Second, there is the need to add a new and unique non-terminal symbol (\texttt {Start\_uni}), which will be the root of the union grammar. This is accomplished by the use of an inductive type definition (\texttt {g\_uni\_nt}), in contrast with the previous case, that used a simple non inductive definition.

The functions \texttt {g\_uni\_sf\_lift\_left} and \texttt {g\_uni\_sf\_lift\_right} simply map sentential forms from, respectively, the first or the second grammar in a pair, and produce sentential forms for the union grammar. This will be useful when defining the rules of the union grammar.

The rules of the union grammar are represented by the inductive definition \texttt {g\_uni\_rules}. Constructors \texttt {Start1\_uni} and \texttt {Start2\_uni} state that two new rules are added to the union grammar: respectively the rule that maps the new root to the root of the first grammar, and the rule that maps the new root to the root of the second grammar. Then, constructors \texttt {Lift1\_uni} and \texttt {Lift2\_uni} simply map rules in first (resp. second) grammar in rules of the union grammar.

Finally, \texttt {g\_uni} describes how to create a union grammar from two arbitrary source grammars. It uses the previous definitions to give values to each of the components of a new grammar definition.

Similar definitions were created to represent the concatenation of any two grammars and the closure of a grammar.

\subsection {Concatenation}

\begin{verbatim}
Definition g_cat_t (g1 g2: cfg): Type:= 
(terminal g1 + terminal g2)%type.

Inductive g_cat_nt (g1 g2: cfg): Type :=
| Start_cat : g_cat_nt g1 g2
| Transf1_cat : non_terminal g1 -> g_cat_nt g1 g2
| Transf2_cat : non_terminal g2 -> g_cat_nt g1 g2.

Definition g_cat_sf_lift_left (g1 g2: cfg)
(c: non_terminal g1 + terminal g1): g_cat_nt g1 g2 + g_cat_t g1 g2:=
  match c with
  | inl nt => inl (Transf1_cat g1 g2 nt)
  | inr t  => inr (inl t)
  end.

Definition g_cat_sf_lift_right (g1 g2: cfg)
(c: non_terminal g2 + terminal g2): g_cat_nt g1 g2 + g_cat_t g1 g2:=
  match c with
  | inl nt => inl (Transf2_cat g1 g2 nt)
  | inr t  => inr (inr t)
  end.

Inductive g_cat_rules (g1 g2: cfg): g_cat_nt g1 g2 -> 
list (g_cat_nt g1 g2 + g_cat_t g1 g2) -> Prop :=
| New_cat: g_cat_rules g1 g2 (Start_cat g1 g2) 
           ([inl (Transf1_cat g1 g2 (start_symbol g1))]++
           [inl (Transf2_cat g1 g2 (start_symbol g2))])%list
| Lift1_cat: forall nt s,
             rules g1 nt s ->
             g_cat_rules g1 g2 (Transf1_cat g1 g2 nt)
             (map (g_cat_sf_lift_left g1 g2) s)
| Lift2_cat: forall nt s,
             rules g2 nt s ->
             g_cat_rules g1 g2 (Transf2_cat g1 g2 nt)
             (map (g_cat_sf_lift_right g1 g2) s).

Definition g_cat (g1 g2: cfg): cfg := {|
non_terminal:= g_cat_nt g1 g2;
terminal:= g_cat_t g1 g2;
start_symbol:= Start_cat g1 g2;
rules:= g_cat_rules g1 g2
|}.
\end{verbatim}

\subsection {Closure}

\begin{verbatim}
Definition g_clo_t (g: cfg): Type:= 
(terminal g)%type.

Inductive g_clo_nt (g: cfg): Type :=
| Start_clo : g_clo_nt g
| Transf_clo : non_terminal g -> g_clo_nt g.

Definition g_clo_sf_lift (g: cfg)
(c: non_terminal g + terminal g): g_clo_nt g + g_clo_t g:=
  match c with
  | inl nt => inl (Transf_clo g nt)
  | inr t  => inr t
  end.

Inductive g_clo_rules (g: cfg): g_clo_nt g -> 
list (g_clo_nt g + g_clo_t g) -> Prop :=
| New1_clo: g_clo_rules g (Start_clo g) 
            ([inl (Start_clo g)]++
             [inl (Transf_clo g (start_symbol g))])
| New2_clo: g_clo_rules g (Start_clo g) []
| Lift_clo: forall nt s,
            rules g nt s ->
            g_clo_rules g (Transf_clo g nt) (map (g_clo_sf_lift g) s).

Definition g_clo (g: cfg): cfg := {|
non_terminal:= g_clo_nt g;
terminal:= g_clo_t g;
start_symbol:= Start_clo g;
rules:= g_clo_rules g
|}.
\end{verbatim}

\subsection {Correctness}

Although simple in their structure, it must be proved that the definitions \texttt {g\_uni}, \texttt {g\_cat} and \texttt {g\_clo} always produce the correct result. In other words, the algorithms embedded in these definitions must be ``certified''. The process of doing such a certification is called ``program verification'', and is one of the main goals of formalization. In order to accomplish this, we must first state theorems, using first-order logic, that capture the expected semantics of these definitions. Finally, we have to derive proofs of the correctness of these theorems. 

This can be done with a pair of theorems for each definition/algorithm: the first relates the output to the inputs, and the other one does the inverse, providing assumptions about the inputs once an output is generated. This is necessary in order to guarantee that the algorithm does only what one would expect, and no more.

\noindent\emph {Concatenation, direct operation:}
\begin{verbatim}
Theorem g_cat_correct (g1 g2: cfg)(s1: sf g1)(s2: sf g2):
generates g1 s1 /\ generates g2 s2 ->
generates (g_cat g1 g2)
          ((map (g_cat_sf_lift_left g1 g2) s1)++
          (map (g_cat_sf_lift_right g1 g2) s2))%list.
\end{verbatim}

The above theorem, for example, states that if context-free grammars \texttt{g1} and \texttt{g2} generate, respectively, strings \texttt{s1} and \texttt{s2}, then the concatenation of these two grammars, according to the proposed algorithm, generates the concatenation of string \texttt{s1} to string \texttt{s2}. As mentioned before, the above theorem alone does not guarantee that \texttt{g\_cat} will not produce outputs other than the concatenation of its input strings. This idea is captured by the following complementary theorem:

\noindent\emph {Concatenation, inverse operation:}
\begin{verbatim}
Theorem g_cat_correct_inv (g1 g2: cfg)(s: sf (g_cat g1 g2)):
generates (g_cat g1 g2) s ->
exists s1: sf g1, 
exists s2: sf g2,
s =(map (g_cat_sf_lift_left g1 g2) s1)++
   (map (g_cat_sf_lift_right g1 g2) s2) /\
generates g1 s1 /\
generates g2 s2.
\end{verbatim}

The idea here is to express that, if a string is generated by \texttt{g\_cat}, then it must only result from the concatenation of strings generated by the grammars merged by the algorithm. Together, these two theorems represent the semantics of the context-free grammar concatenation operation presented. The same ideas have been applied to the statement and proof of the following theorems, relative to the union and closure operations:

\noindent\emph {Union, direct operation:}
\begin{verbatim}
Theorem g_uni_correct (g1 g2: cfg)(s1: sf g1)(s2: sf g2):
generates g1 s1 \/ generates g2 s2 ->
generates (g_uni g1 g2) (map (g_uni_sf_lift_left g1 g2) s1) \/ 
generates (g_uni g1 g2) (map (g_uni_sf_lift_right g1 g2) s2).
\end{verbatim}

\noindent\emph {Union, inverse operation:}
\begin{verbatim}
Theorem g_uni_correct_inv (g1 g2: cfg)(s: sf (g_uni g1 g2)):
generates (g_uni g1 g2) s ->
(s=[inl (start_symbol (g_uni g1 g2))]) \/
(exists s1: sf g1,
(s=(map (g_uni_sf_lift_left g1 g2) s1) /\ generates g1 s1)) \/
(exists s2: sf g2,
(s=(map (g_uni_sf_lift_right g1 g2) s2) /\ generates g2 s2)).
\end{verbatim}

\noindent\emph {Closure, direct operation:}
\begin{verbatim}
Theorem g_clo_correct (g: cfg)(s: sf g)(s': sf (g_clo g)):
generates (g_clo g) nil /\
(generates (g_clo g) s' /\ generates g s -> 
generates (g_clo g) (s'++ (map (g_clo_sf_lift g)) s)).
\end{verbatim}

\noindent\emph {Closure, inverse operation:}
\begin{verbatim}
Theorem g_clo_correct_inv (g: cfg)(s: sf (g_clo g)):
generates (g_clo g) s -> 
(s=[]) \/
(s=[inl (start_symbol (g_clo g))]) \/
(exists s': sf (g_clo g), 
 exists s'': sf g,
 generates (g_clo g) s' /\ generates g s'' /\ 
 s=s'++map (g_clo_sf_lift g) s'').
\end{verbatim}

The proofs of all the six main theorems have been completed (\texttt {g\_uni\_correct} and \texttt {g\_uni\_correct\_inv} for union, \texttt {g\_cat\_correct} and \texttt {g\_cat\_correct\_inv} for concatenation and \texttt {g\_clo\_correct} and \texttt {g\_clo\_correct\_inv} for closure). 

\section {Final remarks}

As an interesting side result, some useful and generic lemmas have also been proved during this process. Among these, for example, one that asserts the context-free characteristic of these derivations:

\begin{verbatim}
Theorem derives_context_free_add:
forall g:cfg,
forall s1 s2 s s': sf g,
derives g s1 s2 -> derives g (s++s1++s') (s++s2++s').
\end{verbatim}

\noindent
and one that states the transitivity of the \texttt {derives} relation:

\begin{verbatim}
Theorem derives_trans:
forall g: cfg,
forall s1 s2 s3: sf g,
derives g s1 s2 ->
derives g s2 s3 ->
derives g s1 s3.
\end{verbatim}

The formalization of this section comprises ~ 1,400 lines of Coq script, and is available for download at: \\
\url {https://github.com/mvmramos/cfg}. \\
\texttt{String} and \texttt{List}.

\chapter {Simplification}

The results reported next are related to the elimination of symbols (terminals and non-terminals) in context-free grammars that do not contribute to the language being generated, and also to the elimination of unit and empty rules, in order to shorten the derivation of the sentences of the language.

\section {Basic definitions}
\label {sec-basic}
Context-free grammars were represented in Coq very closely to the usual algebraic definition $G=(V,\Sigma,P,S)$, where $\Sigma$ is the set of terminal symbols (used in the construction of the sentences of the language generated by the grammar), $N=V-\Sigma$ is the set of non-terminal symbols (representing different sentence abstractions), $P$ is the set of rules and $S \in N$ is the start symbol (also called initial or root symbol). Rules have the form $\alpha \rightarrow \beta$, with $\alpha \in N$ and $\beta \in V^*$.

Basic definitions in Coq are presented below. The $N$ and $\Sigma$ sets are represented separately from $G$ (respectively by types \texttt {non\_terminal} and \texttt {terminal}). Notations \texttt {sf} (sentential form) and \texttt {sentence} represent lists, possibly empty, of respectively terminal and non-terminal symbols and terminal only symbols.

\begin{verbatim}
Variables non_terminal terminal: Type.
Notation sf := (list (non_terminal + terminal)).
Notation sentence := (list terminal).
Notation nlist:= (list non_terminal).
\end{verbatim}

The record representation \texttt {cfg} has been used for $G$. The definition states that \texttt {cfg} is a new type and contains three components. The first is the \texttt {start\_symbol} of the grammar (a non-terminal symbol) and the second is \texttt {rules}, that represent the rules of the grammar. Rules are propositions (represented in Coq by \texttt {Prop}) that take as arguments a non-terminal symbol and a (possibly empty) list of non-terminal and terminal symbols (corresponding, respectively, to the left and right-hand side of a rule). The predicate \texttt {rules\_finite} assures that the set of rules is finite by proving that every rule has a finite length and is built from a finite set of non-terminal symbols and a finite set of terminal symbols (which in turn are represented by the respective lists of symbols).

\begin{verbatim}
Record cfg: Type:= {
start_symbol: non_terminal;
rules: non_terminal -> sf -> Prop;
rules_finite: exists n: nat,
              exists ntl: nlist,
              exists tl: tlist,
              In start_symbol ntl /\
              forall left: non_terminal,
              forall right: sf,
              rules left right ->
              (length right <= n) /\
              (In left ntl) /\
              (forall s: non_terminal, In (inl s) right -> In s ntl) /\
              (forall s: terminal, In (inr s) right -> In s tl)
}.
\end{verbatim}

The example below represents grammar $$G=(\{S,A,B,a,b\},\{a,b\},\{S \rightarrow aS, S \rightarrow b\},S)$$ that generates language $a^*b$:

\begin{verbatim}
Inductive non_terminal: Type:=
| S
| A
| B.

Inductive terminal: Type:=
| a
| b.

Inductive rs: non_terminal -> sf -> Prop:=
  r1: rs S [inr a; inl S]
| r2: rs S [inr b].

Lemma g_finite:
exists n: nat,
exists ntl: nlist,
exists tl: tlist,
In S' ntl /\
forall left: non_terminal,
forall right: sf,
rs left right ->
(length right <= n) /\
(In left ntl) /\
(forall s: non_terminal, In (inl s) right -> In s ntl) /\
(forall s: terminal, In (inr s) right -> In s tl).
Proof.
exists 2. 
exists [S'].
exists [a; b].
split.
- simpl; left; reflexivity.
- intros left right H.
  inversion H.
  + simpl. 
    split.
    * omega.
    * {
      split.
      - left; reflexivity.
      - split.
        + intros s H2.
          destruct H2 as [H2 | H2].
          * inversion H2.
          * {
            destruct H2 as [H2 | H2].
            - left; inversion H2.
              reflexivity.
            - contradiction.
            }
        + intros s H2.
          destruct H2 as [H2 | H2].
          * inversion H2.
            left; reflexivity.
          * {
            destruct H2 as [H2 | H2].
            - inversion H2.
            - contradiction.
            }
      }
  + simpl.
    split.
    * omega.
    * {
      split.
      - left; reflexivity.
      - split.
        + intros s H2.  
          destruct H2 as [H2 | H2].
          * inversion H2.
          * contradiction.
        + intros s H2.
          destruct H2 as [H2 | H2].
          * inversion H2.
            right; left; reflexivity.
          * contradiction.
      }
Qed.

Definition g: cfg _ _:= {|
start_symbol:= S'; 
rules:= rs;
rules_finite:= g_finite
|}.
\end{verbatim}

Another fundamental concept used in this formalization is the idea of \emph {derivation}: a grammar \texttt{g} \emph {derives} a string \texttt {s2} from a string \texttt {s1} if there exists a series of rules in \texttt {g} that, when applied to \texttt {s1}, eventually result in \texttt {s2}. An inductive predicate definition of this concept in Coq (\texttt {derives}) uses two constructors.

\begin{verbatim}
Inductive derives (g: cfg): sf g -> sf g -> Prop :=
| derives_refl: forall s: sf g,
                derives g s s
| derives_step: forall s1 s2 s3: sf g,
                forall left: non_terminal g,
                forall right: sf g,
                derives g s1 (s2 ++ inl left :: s3) ->
                rules g left right ->
                derives g s1 (s2 ++ right ++ s3).
\end{verbatim}

The constructors of this definition (\texttt {derives\_refl} and \texttt {derives\_step}) are the axioms of our theory. Constructor \texttt {derives\_refl} asserts that every sentential form \texttt {s} can be derived from \texttt {s} itself. Constructor \texttt {derives\_step} states that if a sentential form that contains the left-hand side of a rule is derived by a grammar, then the grammar derives the sentential form with the left-hand side replaced by the right-hand side of the same rule. This case corresponds to the application of a rule in a direct derivation step.

A grammar \texttt {generates} a string if this string can be derived from its root symbol. Finally, a grammar \texttt {produces} a sentence if it can be generated from its root symbol.

\begin{verbatim}
Definition generates (g: cfg) (s: sf g): Prop:=
derives g [inl (start_symbol g)] s.

Definition produces (g: cfg) (s: sentence): Prop:=
generates g (map terminal_lift s).
\end{verbatim} 

Function \texttt {terminal\_lift} converts a terminal symbol into an ordered pair of type \texttt {(non\_terminal + terminal)}. With these definitions, it has been possible to prove various lemmas about grammars and derivations, and also operations on grammars, all of which were useful when proving the main theorems of this article. 

The proof that grammar $G$ produces string $aab$ (that is, that $aab \in L(G)$) is given below:

\begin{verbatim}
Lemma G_produces_aab:
produces G [a; a; b].
Proof.
unfold produces.
unfold generates.
simpl.
apply derives_step with (s2:=[inr a;inr a])(left:=S)(right:=[inr b]).
apply derives_step with (s2:=[inr a])(left:=S)(right:=[inr a;inl S]).
apply derives_start with (left:=S)(right:=[inr a;inl S]).
apply r1.
apply r1.
apply r2.
Qed.
\end{verbatim} 

This proof relates directly to the derivations in $S \Rightarrow aS \Rightarrow aaS \Rightarrow aab$, however in reverse order because of the way that \texttt {derives} was defined.

\section {Formalization}
\label {sec-simplification}

The definition of a context-free grammar allows for the inclusion of symbols and rules that might not contribute to the language being generated. Also, context-free grammars might also contain sets of rules that can be substituted by equivalent smaller and simpler sets of rules. Unit rules, for example, do not expand sentential forms (instead, they just rename the symbols in them) and empty rules can cause them to contract. Although the appropriate use of these features can be important for human communication in some situations, this is not the general case, since it leads to grammars that have more symbols and rules than necessary, making difficult its comprehension and manipulation. Thus, simplification is an important operation on context-free grammars.

Let $G$ be a context-free grammar, $L(G)$ the language generated by this grammar and $\epsilon$ the empty string. Different authors use different terminology when presenting simplification results for context-free grammars. In what follows, we adopt the terminology and definitions of \cite {sudkamp-2006}. 

Context-free grammar simplification comprises four kinds of objects, whose definitions and results are presented below:

\begin {enumerate}
\item \label {empty}
An \emph {empty rule} $r \in P$ is a rule whose right-hand side $\beta$ is empty (e.g. $X \rightarrow \epsilon$). We formalize that for all $G$, there exists $G'$ such that $L(G)=L(G')$ and $G'$ has no empty rules, except for a single rule $S \rightarrow \epsilon$ if $\epsilon \in L(G)$; in this case, $S$ (the initial symbol of $G'$) does not appear in the right-hand side of any rule in $G'$;

\item \label {unit} 
An \emph {unit rule} $r \in P$ is a rule whose right-hand side $\beta$ contains a single non-terminal symbol (e.g. $X \rightarrow Y$). We formalize that for all $G$, there exists $G'$ such that $L(G)=L(G')$ and $G'$ has no unit rules;

\item \label {useless} 
$s \in V$ is \emph {useful} (\cite {sudkamp-2006}, p. 116) if it is possible to derive a string of terminal symbols from it using the rules of the grammar. Otherwise $s$ is called an \emph {useless symbol}. A useful symbol $s$ is one such that $s \Rightarrow^* \omega$, with $\omega \in \Sigma^*$. Naturally, this definition concerns mainly non-terminals, as terminals are trivially useful. We formalize that, for all $G$ such that $L(G) \neq \emptyset$, there exists $G'$ such that $L(G)=L(G')$ and $G'$ has no useless symbols;

\item \label {inaccessible} 
$s \in V$ is \emph {accessible} (\cite {sudkamp-2006}, p. 119) if it is part of at least one string generated from the root symbol of the grammar. Otherwise it is called an \emph {inaccessible symbol}. An accessible symbol $s$ is one such that $S \Rightarrow^* \alpha s\beta$, with $\alpha, \beta \in V^*$. We formalize that for all $G$, there exists $G'$ such that $L(G)=L(G')$ and $G'$ has no inaccessible symbols.
\end {enumerate}

Finally, we formalize a unification result: that forall $G$, if $G$ is non-empty, then there exists $G'$ such that $L(G)=L(G')$ and $G'$ has no empty rules (except for one, if $G$ generates the empty string), no unit rules, no useless symbols and no inaccessible symbols.

In all these four cases and five grammars that are discussed next (namely \texttt {g\_emp}, \texttt {g\_emp'}, \texttt {g\_unit}, \texttt {g\_use} and \texttt {g\_acc}), the proof of the predicate \texttt {rules\_finite} is based in the proof of the correspondent predicate for the argument grammar. Thus, all new grammars satisfy the \texttt {cfg} specification and are finite as well.

\subsection {Empty rules}
Result (\ref {empty}) was achieved in two steps. First, the idea of a \emph {nullable} symbol was represented by the definition \texttt {empty}. With it, a new grammar \texttt {g\_emp g} has been created, such that the laguage generated by it matches the language generated by the original grammar (\texttt {g}), except for the empty string. \texttt {g\_emp\_rules} states that every non-empty rule from \texttt {g} is also a rule from \texttt {g\_emp g}, and also adds new rules to \texttt {g\_emp g} where every possible combination of nullable non-terminal symbols that appears in the right-hand side of a rule of \texttt {g} is removed, as long as the resulting righ-hand side is not empty.

\begin{verbatim}
Definition empty 
(g: cfg terminal _) (s: non_terminal + terminal): Prop:=
derives g [s] [].

Inductive g_emp_rules (g: cfg _ _): non_terminal -> sf -> Prop :=
| Lift_direct : 
       forall left: non_terminal,
       forall right: sf,
       right <> [] -> rules g left right ->
       g_emp_rules g left right
| Lift_indirect:
       forall left: non_terminal,
       forall right: sf,
       g_emp_rules g left right ->
       forall s1 s2: sf, 
       forall s: non_terminal,
       right = s1 ++ (inl s) :: s2 ->
       empty g (inl s) ->
       s1 ++ s2 <> [] ->
       g_emp_rules g left (s1 ++ s2).

Definition g_emp (g: cfg _ _): cfg _ _ := {|
start_symbol:= start_symbol g;
rules:= g_emp_rules g;
rules_finite:= g_emp_finite g |}.
\end{verbatim}

Suppose, for example, that $S, A, B, C$ are non-terminals, of which $A, B$ and $C$ are nullable, $a, b$ and $c$ are terminals and $X \rightarrow aAbBcC$ is a rule of \texttt {g}. Then, the above definitions assert that $X \rightarrow aAbBcC$ is a rule of \texttt {g\_emp g}, and also:

\begin {itemize}
\item $X \rightarrow aAbBc$;
\item $X \rightarrow abBcC$;
\item $X \rightarrow aAbcC$;
\item $X \rightarrow aAbc$;
\item $X \rightarrow abBc$;
\item $X \rightarrow abcC$;
\item $X \rightarrow abc$.
\end {itemize}

Grammar \texttt {g\_emp g} has the same initial symbol as \texttt {g} and does not generate the empty string. The second step was to define \texttt {g\_emp' g}, such that \texttt {g\_emp' g} generates the empty string if and only if \texttt {g} generates the empty string. This was done by stating that every rule from \texttt {g\_emp g} is also a rule of \texttt {g\_emp' g} and by adding a new start symbol to \texttt {g\_emp' g} (\texttt {New\_ss}), as well as two new rules that allow \texttt {g\_emp' g} to generate the same set of sentences of \texttt {g}, plus the empty string. 

Notation \texttt {sf'} represents a sentential form built with symbols from \texttt {non\_terminal'} and \texttt {terminal}. Definition \texttt {symbol\_lift} maps a pair of type \texttt {(non\_terminal + terminal)} into a pair of type \texttt {(non\_terminal' + terminal)} by replacing each \texttt {non\_terminal} with the corresponding \texttt {non\_terminal'}.

\begin{verbatim}
Inductive non_terminal': Type:=
| Lift_nt: non_terminal -> non_terminal'
| New_ss.

Notation sf' := (list (non_terminal' + terminal)).

Definition symbol_lift 
(s: non_terminal + terminal): non_terminal' + terminal:=
match s with
| inr t => inr t
| inl n => inl (Lift_nt n)
end.

Inductive g_emp'_rules (g: cfg terminal non_terminal): 
non_terminal' -> sf' -> Prop :=
| Lift_all:
       forall left: non_terminal,
       forall right: sf,
       rules (g_emp g) left right ->
       g_emp'_rules g (Lift_nt left) (map symbol_lift right)
| Lift_empty:
       empty g (inl (start_symbol (g_emp g))) -> 
	   g_emp'_rules g New_ss []
| Lift_start:
       g_emp'_rules g New_ss [inl (Lift_nt (start_symbol (g_emp g)))].

Definition g_emp' (g: cfg non_terminal terminal): 
cfg non_terminal' terminal := {|
start_symbol:= New_ss;
rules:= g_emp'_rules g;
rules_finite:= g_emp'_finite g |}.
\end{verbatim}

Note that the generation of the empty string by \texttt {g\_emp' g} depends on \texttt {g} generating the empty string. 

The proof of the correctness of these definitions is achieved through the following theorem:

\begin{verbatim}
Theorem g_emp'_correct: 
forall g: cfg _ _,
g_equiv (g_emp' g) g /\
(generates_empty g -> has_one_empty_rule (g_emp' g)) /\ 
(~ generates_empty g -> has_no_empty_rules (g_emp' g)).
\end{verbatim}

Two auxiliary predicates have been used: \texttt {has\_one\_empty\_rule} states that the grammar that it takes as its argument has an empty rule whose left-hand side is the initial symbol, and also that all other rules of the grammar are not empty. \texttt {has\_no\_empty\_rules} states that the grammar contains no empty rules at all.

\subsection {Unit rules}

For result (\ref {unit}), definition \texttt {unit} expresses the relation between any two non-terminal symbols $X$ and $Y$, and is true when $X \Rightarrow^* Y$.

\begin{verbatim}
Inductive unit (g: cfg terminal non_terminal) (a: non_terminal): 
non_terminal -> Prop:=
| unit_rule: forall (b: non_terminal),
             rules g a [inl b] -> unit g a b
| unit_trans: forall b c: non_terminal,
              unit g a b ->
              unit g b c ->
              unit g a c.
\end{verbatim}

Grammar \texttt {g\_unit g} represents the grammar whose unit rules have been substituted by equivalent ones. The idea is that \texttt {g\_unit g} has all non-unit rules of \texttt {g}, plus new rules that are created by anticipating the possibile application of unit rules in \texttt {g}, as informed by \texttt {g\_unit}.

\begin{verbatim}
Inductive g_unit_rules (g: cfg _ _): non_terminal -> sf -> Prop :=
| Lift_direct' : 
       forall left: non_terminal,
       forall right: sf,
       (forall r: non_terminal,
       right <> [inl r]) -> rules g left right ->
       g_unit_rules g left right
| Lift_indirect':
       forall a b: non_terminal,
       unit g a b ->
       forall right: sf,
       rules g b right ->  
       (forall c: non_terminal,
       right <> [inl c]) -> 
       g_unit_rules g a right.

Definition g_unit (g: cfg _ _): cfg _ _ := {|
start_symbol:= start_symbol g;
rules:= g_unit_rules g;
rules_finite:= g_unit_finite g |}.
\end{verbatim}

Finally, the correcteness of \texttt {g\_unit} comes from the following theorem:

\begin{verbatim}
Theorem g_unit_correct: 
forall g: cfg _ _,
g_equiv (g_unit g) g /\
has_no_unit_rules (g_unit g).
\end{verbatim}

\subsection {Useless symbols}
For result (\ref {useless}), the idea of an useful symbol is captured by the definition \texttt {useful}.

\begin{verbatim}
Definition useful (g: cfg _ _) (s: non_terminal + terminal): Prop:=
match s with
| inr t => True
| inl n => exists s: sentence, derives g [inl n] (map term_lift s)
end.
\end{verbatim}

The removal of useless symbols comprises, first, the identification of useless symbols in the grammar and, second, the elimination of the rules that use them. Definition \texttt {g\_use\_rules} selects, from the original grammar, only the rules that do not contain useless symbols. The new grammar, without useless symbols, can then be defined as in \texttt {g\_use}.

\begin{verbatim}
Inductive g_use_rules (g: cfg): non_terminal -> sf -> Prop :=
| Lift_use : forall left: non_terminal,
             forall right: sf,
             rules g left right ->
             useful g (inl left) ->
             (forall s: non_terminal + terminal, In s right -> 
             useful g s) -> g_use_rules g left right.

Definition g_use (g: cfg _ _): cfg _ _:= {|
start_symbol:= start_symbol g;
rules:= g_use_rules g;
rules_finite:= g_use_finite g |}.			 
\end{verbatim}

The \texttt {g\_use} definition, of course, can only be used if the language generated by the original grammar is not empty, that is, if the root symbol of the original grammar is useful. If it were useless then it would be impossible to assign a root to the grammar and the language would be empty. The correctness of the useless symbol elimination operation can be certified by proving theorem \texttt {g\_use\_correct}, which states that every context-free grammar whose root symbol is useful generates a language that can also be generated by an equivalent context-free grammar whose symbols are all useful.

\begin{verbatim}
Theorem g_use_correct: 
forall g: cfg _ _,
non_empty g ->
g_equiv (g_use g) g /\
has_no_useless_symbols (g_use g). 
\end{verbatim}

The predicates (a) \texttt {g\_non\_empty}, (b) \texttt {g\_equiv} and (c) \texttt {has\_no\_useless\_symbols} used above assert, respectively, that (a) grammar \texttt {g} generates a language that contains at least one string (which in turn may or may not be empty); (b) two context-free grammars generate the same language and (c) the grammar has no useless symbols at all.

\subsection {Inaccessible symbols}

Result (\ref {inaccessible}) is similar to the previous case, and definition \texttt {accessible} has been used to represent accessible symbols in context-free grammars.

\begin{verbatim}
Definition accessible 
(g: cfg _ _) (s: non_terminal + terminal): Prop:=
exists s1 s2: sf, derives g [inl (start_symbol g)] (s1++s::s2).
\end{verbatim}

Definition \texttt {g\_acc\_rules} selects, from the original grammar, only the rules that do not contain inaccessible symbols. Definition \texttt {g\_acc} represents a grammar whose inacessible symbols have been removed.

\begin{verbatim}
Inductive g_acc_rules (g: cfg): non_terminal -> sf -> Prop :=
| Lift_acc : forall left: non_terminal,
             forall right: sf,
             rules g left right -> accessible g (inl left) -> 
             g_acc_rules g left right.

Definition g_acc (g: cfg _ _): cfg _ _ := {|
start_symbol:= start_symbol g;
rules:= g_acc_rules g;
rules_finite:= g_acc_finite g |}.
\end{verbatim}

The correctness of the inaccessible symbol elimination operation can be certified by proving theorem \texttt {g\_acc\_correct}, which states that every context-free grammar generates a language that can also be generated by an equivalent context-free grammar whose symbols are all accessible.

\begin{verbatim}
Theorem g_acc_correct: 
forall g: cfg _ _,
g_equiv (g_acc g) g /\
has_no_inaccessible_symbols (g_acc g).  
\end{verbatim}

In a way similar to \texttt {has\_no\_useless\_symbols}, the absence of inaccessible symbols in a grammar is expressed by definition \texttt {has\_no\_inaccessible\_symbols} used above.

\subsection {Unification}

If one wants to obtain a new grammar simultaneously free of empty and unit rules, and of useless and inaccessible symbols, it is not enough to consider the previous independent results. On the other hand, it is necessary to establish a suitable order to apply these simplifications, in order to guarantee that the final result satisfies all desired conditions. Then, it is necessary to prove that the claims do hold.

For the order, we should start with (i) the elimination of empty rules, followed by (ii) the elimination of unit rules. The reason for this is that (i) might introduce new unit rules in the grammar, and (ii) will surely not introduce empty rules, as long as original grammar is free of them (except for $S \rightarrow \epsilon$, in which case $S$, the initial symbol of the grammar, must not appear in the right-hand side of any rule). Then, elimination of useless and inaccessible symbols (in either order) in the right thing to do, since they only remove rules from the original grammar (which is specially important because they do not introduce new empty or unit rules).

The formalization of this result is captured in the following theorem, which represents the main result of this work:

\begin{verbatim}
Theorem g_simpl:
forall g: cfg non_terminal terminal,
non_empty g ->
exists g': cfg (non_terminal' non_terminal) terminal,
g_equiv g' g /\
has_no_inaccessible_symbols g' /\
has_no_useless_symbols g' /\
(generates_empty g -> has_one_empty_rule g') /\ 
(~ generates_empty g -> has_no_empty_rules g') /\
has_no_unit_rules g'.
\end{verbatim}

Hypothesis \texttt {non\_empty g} is necessary in order to allow the elimination of useless symbols.

The proof of \texttt {g\_simpl} demands auxilary lemmas to prove that the characteristics of the initial transformations are preserved by the following ones. For example, that all of the unit rules elimination, useless symbol elimination and inaccessbile symbol elimination operations preserve the characteristics of the empty rules elimination operation.

\section {Final remarks}
The formalization of this section comprises ~ 10,000 lines of Coq script, and is available for download at: \\
\url {https://github.com/mvmramos/simplification}. \\
It includes an extensive library of fundamental lemmas and theorems on context-free grammars and derivations.

\chapter {Related Work}
\label {cha-related}

One important initiative is the work of Jean-Christophe Filli\^atre and Judic\"ael Courant \cite {filliatre-1997a} \cite {filliatre-1997b}, whose libraries developed during the mid-1990s are still available as a user contribution under the label ``Computer Science/Formal Languages Theory and Automata'' in the Coq official page in the Internet \cite {coq-contrib}.

As the authors explain, it ``formalises the beginning of formal language theory'' and the main results are restricted to proofs of (i) every rational language being accepted by a finite automata; (ii) every context-free language being accepted by a stack automata; (iii) closure properties for context-free languages (e.g. union). Also, some code extraction of a parser generator from a stack automaton is presented. The whole development consists of 29 files, with approximately 11,000 lines of Coq scripts. As confirmed by one of the authors in a recent personal correspondence (2013), no further development has happened since then. 

The paper by Almeida, Moreira, Pereira and Sousa \cite {almeida-2010b} gives an account of this work and mentions the work of Sébastien Briais as being, somehow, a further development of it. As they explain:

\begin {quote}
``Formalization of finite automata in Coq was first approached by J.-C. Filliâtre... The author’s aim was to prove the pumping lemma for regular languages and the extraction of an OCaml program. More recently, S. Briais ... developed a new formalization of formal languages in Coq, which covers Filliâtre’s work. This formalization includes Thompson construction of an automaton from a re and a naïve construction of two automata equivalence based in testing if the difference of their languages is the empty language.''
\end {quote}

An extensive search, however, did not reveal any paper of publication documenting this work, or any consequences of it, except for a Briais' personal web page where the whole development, consisting of some 9,000 lines of Coq script distributed in 5 files, can be downloaded \cite {briais-2008}.

Moreira and Sousa \cite {moreira-2009} have also published on the formalization of regular languages as a Kleene algebra, using the Coq proof assistant, and point out that their work is partially based on the work of Filliâtre \cite {filliatre-1997a}.

Also available as a user contribution in the Coq official page, the library of Takashi Miyamoto \cite {miyamoto} provides support for the use of regular expressions and expresses the most important results and properties related to it. Spread in 8 files, the project is based on Janusz Brzozowski's work published in \cite {brzozowski-1964}, satisfies the axioms of Kleene algebra (associativity, commutativity, distributivity and identity elements) and comprises some 1,500 lines of Coq code. The work, however, does not map regular expressions to equivalent finite automata. As in the previous case, there is no knowledge of any paper or report on the project, nor any account of the period when this development occurred.

In the end of the 1990s, Constable, Jackson, Naumov and Uribe, from Cornell University, worked in the area of automata theory formalization using the Nuprl proof assistant \cite {constable-1997}. They followed closely the structure of the first textbook by Hopcroft and Ullman \cite {hopcroft-1969} in their development, and as a result they were able to formalize the concepts of symbol, sentence, language and finite automata. The main result presented in the paper is, however, restricted to the formalization of the Myhill-Nerode Theorem and its state minimization corollary. Yet, the authors consider their work to have an exploratory character, as they wanted to investigate ways of formalizing ``computational mathematics'' as opposed to the formalization of ``classical mathematics'', which is based on set theory. Essentially, they express that they ``... want to show that constructive proofs can be used to synthesize programs'' and ``... want to examine whether constructive type theory is a natural expression of the basic ideas of computational mathematics in the same sense that set theory is for purely classical mathematics''. The paper follows closely the definitions of the book, but points out the need for some modifications in order to ``enable a constructive proof''. Also, some gaps in it were corrected and filled in as a consequence of the detailed formulation and checking done by the proof assistant. As a result, not only the theorems are successfully formalized and proved, but also a state minimization algorithm is automatically obtained. Finally, the authors make considerations on the work needed to formalize the other chapters of the book, and express their belief that, with the experience acquired, the whole contents of it could be formalized by a four-person team in less than 18 months.

A similar initiative, from the mid-1990s, is the work of Kaloper and Rudnicki from the University of Alberta \cite {kaloper-1996} They used the Mizar proof assistant to formalize finite automata and the minimization theorem. Also, they formalized Mealy and Moore finite transducers and proved the equivalence of these models. All the work is based on the textbook by Denning, Dennis and Qualitz \cite {denning-1978} and is publicly available at \cite {kaloper-1994}. As pointed out by the authors, some errors in the book were discovered during the formalization process.

The work by Christoph Kreitz on the formalization of deterministic finite automata using the Nuprl proof assistant dates from a decade earlier (1986) and is probably a pioneer in this area \cite {kreitz-1986}. Kreitz used his formalization to state and prove the Pumping Lemma for regular languages and also describe how it could be extended to represent non-deterministic finite automata and automata with empty transitions. Code extracted from the proof allows the user to compute the number $n$ of the Lemma and, given a string $w$, $|w|\ge n$, split it into substrings $x, y$ and $z$, $w=xyz$, that satisfy the requirements of the Lemma. Their work was based on the 1979 version of Hopcroft and Ullman's textbook \cite {hopcroft-1979}.

A recent and important reference is the work of Christian Doczkal, Jan-Oliver Kaiser and Gert Smolka \cite {doczkal-2013}. Following the structure of the book by Kozen \cite {kozen-1997}, they did a fairly complete formalization of regular languages theory. First of all, they managed to represent regular expressions, deterministic and non-deterministic finite automata, and then to prove the main results of regular language theory, such as the equivalence of these representations, the Myhill-Nerode theorem, the existence and uniqueness of a minimal finite state automata, plus closure properties and the pumping lemma. All the development was done in Coq, is only 1,400 lines long, and benefited from the use of the SSReflect Coq plug-in, which offers direct support for finite structures such as graphs and finite types.

Nelma Moreira and David Pereira, from Universidade do Porto, among other researchers and universities of Portugal, have done extensive formalization and published many papers on related subjects using the Coq proof assistant. These include topics such as the mechanization of Kleene algebra \cite {moreira-2009}, verification of the equivalence of regular expressions \cite {almeida-2010a} \cite {moreira-2012} and the construction of finite automata from regular expressions \cite {almeida-2010b}.

The decision of the equivalence of regular expressions, in particular, has deserved a lot of attention recently, and a variety of approaches have been used by different authors, besides Moreira and Pereira. Thomas Braibant and Damien Pous have developed a Coq tactic for establishing the equality of regular expressions and deciding Kleene algebras \cite {braibant-2010} \cite {braibant-2012}. Andrea Asperti \cite {asperti-2012b} has developed an efficient algorithm for testing regular expression equivalence and also formalized in Matita the equivalence of regular expressions and deterministic finite automata. Thierry Coquand and Vincent Siles \cite {coquand-2011} and Alexander Krauss and Tobias Nipkow \cite {krauss-2012} have recent papers published in this area, respectively using Coq and Isabelle/HOL.

Jacques-Henri Jourdan, François Pottier and Xavier Leroy \cite {jourdan-2012} developed a validator for LR(1) parsers in Coq. It checks that the recognizer fully matches the language generated by the corresponding context-free grammar, and is an important contribution in the construction of certified compilers.

Outside of the Coq universe, Michael Norrish \cite {norrish-2011} has done research on mechanization of some computability theory using $\lambda$-calculus and recursive functions. Xu, Zhang and Urban, more recently, have developed a formal model of a Turing Machine and related it to abacus machines and recursive functions \cite {xu-2013}. Wu, Zhang and Urban worked on a version of the Myhill-Nerode theorem that uses regular expressions instead of finite automata \cite {wu-2011}. Tom Ridge \cite {ridge-2011} developed an automatic parser generator for general context-free grammars. Michael Norrish and Aditi Barthwal have published on the mechanization of the equivalence between pushdown automata and context-free grammars \cite {barthwal-norrish-2010b}, the formalization of normal forms for context-free grammars \cite {barthwal-norrish-2010a} and more recently on the mechanization of some context-free language theory \cite {barthwal-norrish-2013}. Finally, Berghofer and Reiter formalized a library for automata that served as support for a decision procedure for Presburger arithmetic \cite {berghofer-2009}. All these works have been developed with the HOL and Isabelle proof assistants.

Moving up the language hierarchy, Andrea Asperti and Wilmer Ricciotti \cite {asperti-2012a} have used the Matita theorem prover to prove basic results on Turing machines, including the existence of a certified universal Turing machine.

As it seems, with a few exceptions, the interest in the formalization of language and automata theory is recent, fragmented and largely concentrated on regular language theory. Not much has been done so far in order to build a complete framework for language and automata theory in Coq or any other proof assistant. In these works, focus has been given to results associated to specific devices (finite automata, regular expressions, context-free grammars, Turing Machines etc) in an unrelated way in most of the cases. The few initiatives that were aimed at building such a framework (e.g. Filliâtre's and Constable's) apparently did not develop further.

\chapter {Contributions}
\label {cha-contrib}
When the work is complete, it will be the first development that the author is aware of to cover the formalization of language and automata theory in a systematic and complete way, ranging from the most simple to the most complex language classes, considering different representations and the most important results of the whole theory. Also, it should be useful for a few different purposes. Among them, it is worth mentioning:

\begin {itemize}
\item \emph {Make available a complete and mathematically precise description of the objects and the behaviour of the objects of language and automata theory.} \\
This can be very helpful when new developing representations for similar devices and in teaching computation theory classes.
\item \emph {Offer fully checked and mechanized demonstrations of its main results.} \\
This represents the guarantee that the proofs are correct and that the remaining errors in the informal demonstrations, if any, could finally and definitely be corrected. Also, this means that these proofs could easily be reused in different contexts with minor efforts.
\item \emph {Allow for the certified and efficient implementation of its relevant algorithms in a proper programming language.} \\
Deriving efficient algorithms from proof terms is an attractive side effect of formalization, and could enable the construction of a library of routines that could then be used safely for a wide range of applications.
\item \emph {Permit the experimentation in an educational environment in the form of a tool set.} \\
Once the source code of the development is made available, teachers and students could use it to learn and experiment with the objects and concepts of language and automata theory in a software laboratory where further practical observations and developments could be done independently. Also, the material could be deployed as the basis for a course on the theoretical foundations of computing, exploring simultaneously not only language and automata theory, but also logic, proof theory, type theory and models of computation. It could also serve as the basis for a course on formal mathematics, interactive theorem provers and Coq.
\item \emph {Expertise and knowledge.} \\
The essence of formalization should come into light with the accomplishment of this project. This shall enable the application of similar principles to the formalization of other theories, and allow for the multiplication of the knowledge among students and colleagues. Considering the growing interest in formalization in recent years, this project could be considered as a good technical preparation for dealing with the challenges of theory and computer program developments of the future.
\end {itemize}

\chapter {Conclusions}
\label {cha-conclusions}
The present paper reports an ongoing research effort towards formalizing the classical context-free language theory, initially based only on context-free grammars, in the Coq proof assistant. All important objects have already been described and basic closure operations on grammars have already been implemented. Proofs of the correctness of the concatenation, union and closure operations (for both direct and inverse ways) were constructed. Also, different simplification strategies on grammars have been implemented. Proofs of the their correctness were successfully constructed. So far, a considerable amount of work has already been done, mainly of logical (or declarative) character, not computational. 

It is expected that, after completion, this research project will have formalized all or at least a large part of context-free language theory (except for automata), something that has been implemented only partially so far. Regarding the context, this work tries to demonstrate the growing importance of formal developments in both theoretical and technological environments by presenting recent and representative projects. In this sense, the present project reflects the current trend in both mathematics and computer science towards formalization using interactive theorem provers.

Also, it has been shown that the present project has not come out of the blue, but rather finds its place among many others that have been published since the mid-1980s. As discussed, most of these have a narrow focus and are restricted to specific language classes (regular, context-free etc) or devices (regular expressions, finite automata etc). Nevertheless, a considerable amount of work has already been devoted to the subject by different authors, and some of it will hopefully be used to the benefit of the present project. 

When the work is complete, it should be useful for a few different purposes. Among them, to make available a complete and mathematically precise description of the behavior of the objects of context-free language theory. Second, to offer fully checked and mechanized demonstrations of its main results. Third, to provide a library with basic and fundamental lemmas and theorems about context-free grammars and derivations that can be used as a starting point to prove new theorems and increase the amount of formalization for context-free language theory. Fourth, to allow for the certified and efficient implementation of its relevant algorithms in a programming language. Fifth, to permit the experimentation in an educational environment in the form of a tool set, in a laboratory where further practical observations and developments can be done, for the benefit of students, teachers, professionals and researchers.

The authors acknowledge the fruitful discussions and contributions of Nelma Moreira (Departamento de Ciência de Computadores da Faculdade de Ciências da Universidade do Porto, Portugal) and José Carlos Bacelar Almeida (Departamento de Informática da Universidade do Minho, Portugal) to this work.

\bibliographystyle {acm}
\bibliography {arXiv}
\addcontentsline {toc}{chapter}{Bibliography}
\end {document}